\documentclass[%
aip,
amsmath,amssymb,
reprint,%
]{revtex4-2}

\usepackage{graphicx}
\usepackage{dcolumn}
\usepackage{bm}

\usepackage[utf8]{inputenc}
\usepackage{mathptmx}

\newcommand{\Eq}[1]{Eq.\,\ref{#1}}
\newcommand{\Fig}[1]{Fig.\,\ref{#1}}
\newcommand{\Sec}[1]{Sec.\,\ref{#1}}
\newcommand{\Tab}[1]{Table \,\ref{#1}}
\newcommand{\be}{\begin{equation}}
	\newcommand{\ee}{\end{equation}}
\newcommand{\bea}{\begin{eqnarray}}
	\newcommand{\eea}{\end{eqnarray}}
\newcommand{\bd}{\begin{displaymath}}
	\newcommand{\ed}{\end{displaymath}}
\newcommand{\ben}{\begin{enumerate}}
	\newcommand{\een}{\end{enumerate}}

\newcommand{\Onlinecite}[1]{Ref.\,\onlinecite{#1}} 

\newcommand{\nm}{\mathrm{nm}} 
\newcommand{\ext}{\mathrm{ext}} 
\newcommand{\abs}{\mathrm{abs}} 
\newcommand{\sca}{\mathrm{sca}} 
\newcommand{\inc}{\mathrm{inc}} 

\newcommand{\siex}{\ensuremath{\sigma_\ext}}  
\newcommand{\siexsb}{\ensuremath{\hat{\sigma}_\ext}} 
\newcommand{\siexsbl}{\ensuremath{\hat{\sigma}_{\Lambda}}} 

\newcommand{\sisc}{\ensuremath{\sigma_\sca}} 
\newcommand{\siscl}{\ensuremath{\sigma_{\sca,\Lambda}}} 
\newcommand{\sia}{\ensuremath{\sigma_\abs}} 
\newcommand{\sial}{\ensuremath{\sigma_{\abs,\Lambda}}} 

\newcommand{\ON}{\ensuremath{\mathrm{ON}}} 
\newcommand{\NA}{\ensuremath{\mathrm{NA}}} 
\newcommand{\ri}{\ensuremath{r_\mathrm{i}}} 
\newcommand{\rs}{\ensuremath{r_\mathrm{s}}} 
\newcommand{\rh}{\ensuremath{r_\mathrm{h}}} 
\newcommand{\rl}{\ensuremath{r_\mathrm{l}}} 

\newcommand{\Nfw}{\ensuremath{N_\mathrm{fw}}} 
\newcommand{\Na}{\ensuremath{N_\mathrm{a}}} 
\newcommand{\dpix}{\ensuremath{d_\mathrm{px}}} 

\newcommand{\ez}{\ensuremath{\epsilon_0}} 
\newcommand{\ep}{\ensuremath{\epsilon_\mathrm{NP}}} 
\newcommand{\emed}{\ensuremath{\epsilon_\mathrm{m}}} 
\newcommand{\nmed}{\ensuremath{n_\mathrm{m}}} 

\newcommand{\Pa}{\ensuremath{P_\abs}} 
\newcommand{\Ps}{\ensuremath{P_\sca}}  
\newcommand{\Iin}{\ensuremath{I_\inc}} 

\newcommand{\hpol}{\ensuremath{\alpha}} 
\newcommand{\ipol}{\ensuremath{\alpha_j}} 
\newcommand{\Li}{\ensuremath{L_j}} 

\newcommand{\unprimeu}{\ensuremath{\vec{e}_\iota}} 
\newcommand{\primeu}{\ensuremath{\vec{e}_\kappa}} 

\newcommand{\ve}{\ensuremath{\vec{E}}} 
\newcommand{\vp}{\ensuremath{\vec{p}}} 
\newcommand{\hr}{\ensuremath{R}} 
\newcommand{\hrt}{\ensuremath{R^\intercal}} 

\newcommand{\Ni}{\ensuremath{N_\mathrm{i}}} 
\newcommand{\Nr}{\ensuremath{N_\mathrm{r}}} 
\newcommand{\gamp}{\ensuremath{\gamma_\mathrm{P}}} 
\newcommand{\alam}{\ensuremath{\alpha_\Lambda}} 
\newcommand{\siexl}{\ensuremath{\sigma_\Lambda}} 
\newcommand{\siexsl}{\ensuremath{\hat{\sigma}_{\Lambda,\mathrm{m}}}} 
\newcommand{\siexsrl}{\ensuremath{\hat{\sigma}_\Lambda}} 
\newcommand{\siexml}{\ensuremath{\bar{\sigma}_\Lambda}} 
\newcommand{\aisl}{\ensuremath{\hat{\alpha}_{\Lambda,\,\mathrm{m}}}} 
\newcommand{\aiml}{\ensuremath{\bar{\alpha}_\Lambda}} 
\newcommand{\aim}[1]{\ensuremath{\bar{\alpha}_{#1}}} 
\newcommand{\ais}[1]{\ensuremath{\hat{\alpha}_{#1}}} 

\newcommand{\brat}{\ensuremath{b/a}} 
\newcommand{\crat}{\ensuremath{c/a}} 
\newcommand{\cbrat}{\ensuremath{c/b}} 

\newcommand{\omp}{\ensuremath{\omega_\mathrm{p}}}
\newcommand{\ebnd}{\ensuremath{\epsilon^\mathrm{b}}}

\newcommand{\Gz}{\ensuremath{\Gamma_0}}

\newcommand{\vf}{\ensuremath{\nu_\mathrm{F}}}

\newcommand{\Sbar}{\ensuremath{\bar{S}}}

\mathchardef\mhyphen="2D 

\newcommand{\DV}{\ensuremath{D_V}} 

\newcommand{\DP}{\ensuremath{D_\mathrm{p}}} 
\newcommand{\DPb}{\ensuremath{\bar{D}_\mathrm{p}}} 
\newcommand{\DPh}{\ensuremath{\hat{D}_\mathrm{p}}} 

\newcommand{\cone}{\ensuremath{\textcolor{blue}{450\,\nm}}} 
\newcommand{\ctwo}{\ensuremath{\textcolor{cyan}{500\,\nm}}} 
\newcommand{\cthr}{\ensuremath{\textcolor{green}{550\,\nm}}} 
\newcommand{\cfour}{\ensuremath{\textcolor{orange}{600\,\nm}}} 
\newcommand{\cfive}{\ensuremath{\textcolor{red}{650\,\nm}}} 
\newcommand{\csix}{\ensuremath{\textcolor{deepred}{700\,\nm}}} 

\newcommand{\lone}{\ensuremath{\textcolor{blue}{450}}} 
\newcommand{\ltwo}{\ensuremath{\textcolor{cyan}{500}}} 
\newcommand{\lthr}{\ensuremath{\textcolor{green}{550}}} 
\newcommand{\lfour}{\ensuremath{\textcolor{orange}{600}}} 
\newcommand{\lfive}{\ensuremath{\textcolor{red}{650}}} 
\newcommand{\lsix}{\ensuremath{\textcolor{deepred}{700}}} 

\newcommand{\gamr}{\ensuremath{\gamma_\mathrm{R}}} 

\newcommand{\sirad}{\ensuremath{\sigma_{\Lambda} (\gamr)}} 

\newcommand{\delsig}{\ensuremath{\delta_\sigma}} 

\newcommand{\wl}{\ensuremath{w(\lambda)}} 
\newcommand{\cl}{\ensuremath{c(\lambda)}} 
\newcommand{\il}{\ensuremath{l(\lambda)}} 

\newcommand{\sigs}{\ensuremath{\sigma_\mathrm{s}}} 
\newcommand{\sigh}{\ensuremath{\sigma_\mathrm{h}}} 
\newcommand{\sigl}{\ensuremath{\sigma_\mathrm{l}}} 

\newcommand{\als}{\ensuremath{\alpha_\mathrm{s}}} 
\newcommand{\gams}{\ensuremath{\gamma_\mathrm{s}}} 
\newcommand{\Npol}{\ensuremath{N_\mathrm{P}}} 
\newcommand{\halpha}{\ensuremath{\hat{\alpha}}} 
\newcommand{\hgamma}{\ensuremath{\hat{\gamma}}} 

\usepackage{textcomp}
\usepackage[dvipsnames]{xcolor}
\definecolor{blue}{rgb}{0 0 1}
\definecolor{cyan}{rgb}{0 0.92 0.92}
\definecolor{green}{rgb}{0 0.9 0}
\definecolor{orange}{rgb}{1 0.6 0}
\definecolor{red}{rgb}{1 0 0}
\definecolor{deepred}{rgb}{0.7 0 0}

\begin{document}

\preprint{AIP/123-QED}

\title[Sub-nanometer 3D morphometric precision of polarisation-resolved wide-field optical extinction microscopy]{Sub-nanometer 3D morphometric precision of polarisation-resolved wide-field optical extinction microscopy determines the roundness of individual gold nanospheres}

\author{Lukas M Payne}
\affiliation{School of Physics and Astronomy, Cardiff University, The Parade, Cardiff CF24 3AA, United Kingdom}
\affiliation{ 
	School of Biosciences, Cardiff University, Museum Avenue, Cardiff CF10 3AX, United Kingdom
}%

\author{Furqan Alabdullah}
\affiliation{ 
	School of Engineering, Cardiff University, The Parade, Cardiff CF24 3AA, United Kingdom
}%
\affiliation{ 
	School of Biosciences, Cardiff University, Museum Avenue, Cardiff CF10 3AX, United Kingdom
}%
\affiliation{ 
	Engineering Technical College of Al-Najaf, Al-Furat Al-Awsat Technical University, Najaf 31001, Iraq
}%

\author{Paola Borri}
\affiliation{ 
	School of Biosciences, Cardiff University, Museum Avenue, Cardiff CF10 3AX, United Kingdom
}%

\author{Wolfgang Langbein}
\affiliation{Cardiff University School of Physics and Astronomy, The Parade, Cardiff CF24 3AA, United Kingdom}
\affiliation{Corresponding author, langbeinww@cardiff.ac.uk}

\date{\today}

\begin{abstract}
Quantitative polarisation-resolved optical extinction microscopy of individual plasmonic nanoparticles has recently been introduced as a powerful tool to characterise the nanoparticle's morphology with a precision comparable to electron microscopy, while using a simple optical microscope [Nanoscale 12, 16215 (2020)]. Here we provide a step change by adding measurements for  radial polarisation in the condenser back focal plane, probing plasmonic resonances polarised in axial direction. The combined linear and radial polarisation measurements provide a significantly enhanced precision of the retrieved 3D morphology, as we show on defect-free ultra-uniform gold nanospheres of 30\,nm nominal diameter characterised by transmission electron microscopy. The measured cross-sections are modelled for an ellipsoidal particle, determining the three semi-axes and rotation angles by fitting the measurements. The material permittivity and surface damping providing the best fit are found. The  particle aspect ratio is determined with a precision better than 5\%, and the size with an impressive precision of $0.1\,\nm$. Notably, corrections to the Rayleigh-Gans ellipsoid model due to retardation are significant even though the particle diameters are more than an order of magnitude smaller than the wavelength in the medium. Taking them into account improves the shape accuracy.  
\end{abstract}

\maketitle


\section{\label{sec:Intro} Introduction}
Over the last decades, nanoparticles (NPs) and nanostructured materials have become increasingly important in many fields, from biological and medical imaging,\cite{HanNS19,PopeLSA23} to sensing,\cite{KneippACSN17, PallaresNS19} catalysis,\cite{AtwaterNM10, HutchingsACSCS18} and drug-delivery.\cite{SunACIE14, LiuM22} Considering the plethora of nano-objects produced by human activity, studies of NP wastes and environmental contaminations\,\cite{TuranCAC22} have attracted increasing attention. Throughout these spaces, the size and shape of NPs play key roles, hence their accurate characterization is important.\cite{ChaturvediJNR24} 

For NPs larger than the diffraction limit of visible light of around $250$\,nm, optical microscopy can directly resolve NP morphology. However, for smaller NPs such a characterization remains a significant challenge. To date, the industry standard for NP morphometry remains transmission electron microscopy (TEM),\cite{PyrzL08, ChaturvediJNR24} but due to high capital and operating costs, TEM is often not economically viable. Additionally, TEM is time-consuming, specifically if tomography is needed for 3D analysis, limiting the number of NPs measured and thus the statistical relevance of the results. Furthermore, the high energy electron beam used can lead to NP damage/reshaping during measurements. These aspects hinder the applicability of TEM to provide high-throughput and reliable statistical assessment of samples, a critical feature to quality control in NP manufacturing.

Apart from microscopy, in recent years techniques exploiting NP mechanical properties have shown promise in NP sizing, such as centrifugation,\cite{CalzolaiFACA12, CaputoJCR19} field-flow fractionation (FFF) in various implementations,\cite{CaputoJCR19, GioriaNM13, WagnerAC14} and tunable resistive pulse sensing (TRPS).\cite{VogelAC11, HenriquezA04, CaputoJCR19} Various versions are commercially available as benchtop devices, but these techniques also exhibit limitations. Centrifugation methods yield ensemble properties of particles only, and all methods provide size but not shape metrics, except FFF which provides indirect shape information.

Popular optical methods for NP morphometry use the light scattered from an incident field by NPs in suspension, to monitor their motion in a viscous liquid medium. These methods include dynamic light scattering (DLS),\cite{Berne76} nanoparticle tracking analysis (NTA),\cite{BellL12,SilmoreACSN19} and interferometric NTA (iNTA).\cite{KashkanovaNM22} Both DLS and NTA are implemented in commercially available benchtop devices, with DLS being an industrial standard for rapid sample size characterization. DLS is an ensemble measurement technique, strongly affected by sample polydispersity, while NTA and iNTA both provide single particle measurements, overcoming this issue. However, all three assume a spherical particle shape undergoing Brownian motion to determine the hydrodynamic radius, which is in itself distinct from the physical radius. Hence, these methods provide no shape information, and use the optical measurements only indirectly to determine the NP size. A nanofluidic scattering-based technique also exists, but is presently restricted to determining the hydrodynamic radius as well.\cite{SpackovaNM22} Alternatively, depolarised dynamic light scattering (DDLS) leverages detection of the co- and cross-polarised components of the scattering signal to measure the translational and rotational diffusion coefficients, and subsequently the length and diameter of the particles assuming a rod or cylindrical-like particle model. However, being a DLS-based technique, it still lacks individual particle measurements, and the determined values are reflective of the apparent size from the NPs motion through liquid.  Other recent advances in photopolarimetric techniques without\cite{WalkerMST04} and with\cite{DawdaSR24} structured illumination exhibit similar limitations. Additional methods include through-focus scanning optical microscopy (TSOM)\cite{AttotaAPL14} and through-focus interferometric nanoparticle imaging\cite{AvciAO17, TruebIEEE17} which use the measured signal from particles at different focus positions.

Complementary approaches directly measure optical properties for diffraction-limited NP morphometry, using either the complex polarisability,\cite{KhadirO20, GentnerACSN24} or the optical cross-sections, namely absorption \sia,\cite{BoyerS02, HusnikPRL12, TcherniakNL10} scattering \sisc,\cite{CrutCSR14, PayneAPL13, ZilliACSP19} and extinction \siex.\cite{ArbouetPRL04, MuskensPRB08, HusnikPRL12, PayneAPL13} \sia\ and \sisc\ are fundamentally related to the NP polarisability, and quantify the ability of an object to absorb or scatter light, respectively. The extinction cross-section, \siex, is the sum of the absorption and scattering, and is given by the effective area blocked by the particle for a collimated incident field. The polarisability and the cross-sections of NPs of given size and shape can be calculated by analytical models only in special cases, such as spheres,\cite{MieAP08} or ellipsoids within the dipole approximation,\cite{BohrenBook83} while for more complex shapes numerical models are needed.\cite{LobanovPRA18, WangNSA20, WangNS22} These optical methods may be employed in conjunction with a physical model, to compare theoretical predictions of morphometry-dependent optical properties with optical measurements, and then infer the NP size and shape, frequently referred to as `solving the inverse problem'.

Recently, our group established a methodology for 3D morphometry of single NPs called the Optical Nanosizer (\ON).\cite{PayneNS20} As a wide-field microscopy modality, the ON offers simultaneous measurement of up to hundreds of NPs in a single field of view, with a sensitivity down to $\siex\approx1\,$nm$^2$, similar to the cross-section of a gold nanosphere (GNS) of $D\approx2\,$nm diameter. In this method, a series of brightfield transmission images of NPs deposited on a coverslip are captured on a conventional microscope under illumination at different wavelengths and linear polarisations. Dark-field images can be additionally taken to distinguish scattering and absorption cross-sections.\cite{ZilliACSP19} Images are analysed as differential transmission contrast, to quantify \siex\ for each NP in the field of view. By comparing the Rayleigh--Gans model for ellipsoidal particles with the measurements of \siex\ for each NP in the images, we inferred the size and shape of nominally spherical small (10 to 30\,nm diameter) gold nanoparticles, and gold nanorods (10\,nm diameter, 30\,nm length). The precision of the method was found to be $\pm 0.5\,\nm$ for size, and about $10\%$ for the shape anisotropy. In terms of accuracy, we found the measurements of GNS having nominal diameter, $D=30\,$nm, to be within $1\,$nm of TEM characterization on the same sample, i.e. showing only about $4\%$ discrepancy. Notably, for shape retrieval we used linearly polarised excitation light in the condenser back focal plane with varying in-plane angle. Even with a high numerical aperture (NA) condenser focussing the light, only a small fraction of light at the sample is polarised out of plane, i.e. axially. Hence, the measurements are dominated by excitation polarisations in the sample plane, and thus are more sensitive to the in-plane geometry of the NP. The out-of-plane geometry was therefore indirectly retrieved by spectral properties using the dependence of surface plasmon resonance frequencies on particle morphology. 

In this work, we present an ON with 3D shape recovery independent of orientation, by using both linear and radially polarised illumination. In conjunction with a high \NA\ condenser, radially polarised light creates a strong axial polarisation component at the sample. This provides a direct probe of the out-of-plane NP geometry, while the linear in-plane polarised illumination probes the in-plane NP geometry. We examine nominally $D=30\,\nm$ `ultra-uniform' gold nanospheres (UGNS) to demonstrate the ON capabilities. Notably, these UGNS had been characterized with high-angle annular dark-field scanning transmission electron microscopy (HAADF-STEM) in our previous works,\cite{PayneNS20, PayneJCP21} including tomography on selected cases, providing a reliable morphometry ground truth. The new ON method shows a significant improvement in precision and accuracy of 3D size and shape retrieval, reaching impressive sub-nm values. 

\section{Model: Permittivity, polarisability, and optical cross-sections}
\label{sec:perm}

The previous \ON\ analytical developments were described in detail in \Onlinecite{PayneNS20, PayneJCP21}. We build here on this analysis, extending it to add the treatment of the radially polarised illumination. We briefly recap the model in the following. The morphometry of each particle is derived from the measured cross-sections by solving `the inverse problem,' using a theoretical framework relating the geometry of a NP to its optical cross-sections. Modelling the optical cross-sections requires knowledge of the complex permittivity of the NP material. In the literature, different measured spectra of the permittivity of gold are provided. We considered three of these, namely the data of Johnson \& Christy (JC),\cite{JohnsonPRB72} McPeak et al. (MP),\cite{McPeakACSP15} and Olmon et al. (OL).\cite{OlmonPRB12} They were measured by spectroscopic ellipsometry on thin polycrystalline films (JC, MP) or single crystalline gold (OL). To account for a surface damping effect in the NPs, we use the following approach. The permittivity is described by a Drude model with a bound electron contribution stemming from the interband transitions from the $d$ bands into the conduction band, and surface damping effects, given by,\cite{MasiaPRB12, ZilliACSP19}
\be\label{eq:NPepsilon}
\ep=1-\frac{\omp^2}{\omega(\omega+i\Gamma)}+\ebnd(\omega),
\ee
with $\omp$ the plasma frequency, \ebnd\ the bound electron contribution\cite{GuerresiPRB75, MasiaPRB12}, and $\Gamma$ the Drude damping rate. A surface damping term is included in $\Gamma$, which is approximated\cite{VoisinPRB04} by 
\be\label{eq:gam0}
\Gamma=\Gz+g\frac{\vf}{R},
\ee
with the NP radius, $R=D/2$, the bulk damping rate, \Gz, the Fermi velocity in gold,\cite{GallJAP16} $\vf=1.4\times10^6\,$m/s, and a dimensionless number, $g$, parameterizing the surface damping -- Table S.1 of \Onlinecite{ZilliACSP19} provides the complete parameterization including $\ebnd(\omega)$. Assuming $g=0$, we determine the values of all the other parameters by using \Eq{eq:NPepsilon} to independently fit each experimental gold permittivity dataset. Using these parameters, we determine a series of \ep\ evaluated with different values of $g\in[0,4.5]$, covering the range of $g$ reported in the literature.\cite{MuskensPRB08,MasiaPRB12}

We use the Rayleigh--Gans model, describing ellipsoidal NPs much smaller than the light wavelength ($\lambda$), to determine the NP  polarisability using the complex material permittivity and geometry of the NP, and the permittivity, \emed, of the surrounding medium. As intruduced in \Onlinecite{PayneJCP21}, we use a Cartesian reference system of unit vectors $\primeu'$, with $\kappa=x',\,y'$, and $z'$ which respectively point in the directions of the three semi-axes of an ellipsoidal NP of length ($a$, $b$, $c$), which we order such that $a\ge b\ge c$. In this basis, the polarisability tensor $\hpol'$ is a diagonal matrix. The diagonal elements are the polarisabilities for fields oriented along each of the NP's semi-axes, given by  
\be\label{eq:poleps}
\ipol'=V\ez \frac{\ep-\emed}{\emed+\Li(\ep-\emed)},
\ee
where $V$ is the NP volume, \ez\ is the free space permittivity, \emed\ is the dielectric function of the surrounding medium, and  \Li\ are depolarisation factors, with $j\in\{1,\,2,\,3\}$. The geometry of the NP determines \Li. For example, for a sphere, $\Li=1/3$. \emed\ is assumed to be real and frequency independent. All materials are assumed to be non-magnetic, i.e. with relative permeability $\mu_\mathrm{r}=1$. We note that metals exhibit a zero crossing versus wavelength of the real part of the denominator in \Eq{eq:poleps} at $\Re\{\ep\}=(1-L_j^{-1})\emed$, which results in a peak in \siex\ called the localised surface plasmon resonance (LSPR).  For an incident field $\ve'$, the induced electric dipole moment is given by
\be\label{eq:dimom}
\vp'=\hpol'\ve'.
\ee

We choose a Cartesian coordinate system to describe the laboratory frame, with axes $\iota \in \{ x, y, z\}$, and associated unit vectors \unprimeu, where $\vec{e}_z$ points along the optical path, and $\vec{e}_x$, $\vec{e}_y$ span the sample plane. To transform $\hpol'$ into the laboratory reference frame, we define\cite{PayneNS20} a 3D rotation matrix using $\hr=R_\psi R_\theta R_\phi$ and $\hrt = R_\phi^\intercal R_\theta^\intercal R_\psi^\intercal$, with $\phi$, $\theta$, $\psi$ the angles of rotation about $\vec{e}_x$, $\vec{e}_y$, $\vec{e}_z$,  respectively.
We transform a vector from the NP frame into the laboratory frame with $\vec{v}=\hr \vec{v}'$. \Eq{eq:dimom} becomes $\hrt\vp=\hpol'\hrt\ve$, such that the polarisability in the laboratory frame is given by
\be\label{eq:rotpol}
\hpol=\hr\hpol'\hrt.
\ee
The scattering and absorption cross-sections, \sisc\ and \sia\, are the power scattered ($\Ps$) or absorbed ($\Pa$) from the incident field by the NP relative to the intensity, \Iin, of the incident field, $\sisc=\Ps/\Iin$ and $\sia=\Pa/\Iin$.  Within the dipole approximation $D\ll\lambda$ the optical cross-sections are given by
\be\label{eq:abs}
\sia(\ve)=\frac{k}{\ez}\frac{\Im(\ve^*\cdot\vp)}{|\ve|^2}\,,
\ee
and
\be\label{eq:sisc}
\sisc(\ve)=\frac{k^4}{6\pi\ez^2}\frac{|\vp|^2}{|\ve|^2}\,,
\ee
with ($*$) indicating complex conjugation, and $k=2\pi\nmed/\lambda$, the wavenumber in the medium of refractive index $\nmed=\sqrt{\emed}$.

To compare the ellipsoid geometry with plan-view TEM characterisation, we define a mean aspect ratio (MAR), $\sqrt{bc}/a$, and a projected aspect ratio (PAR) given by $b'/a'$, with the semi-axes lengths $a'$ and $b'$ in the $xy$-plane of the ellipsoid projected along the $z$ axis  (for calculation details see \Sec{sec:project}). We also define the corresponding projected diameter as $\DP=2\sqrt{a' b'}$ and a volume-equivalent diameter as $\DV=2\sqrt[3]{a b c}$. 

\section{Experiment}\label{sec:Exp}

A detailed description of the microscope setup and data analysis are provided in \Onlinecite{PayneNS20}. A brief overview of the set-up and samples used in this work is given in the following.

\subsection{Optical Measurements} \label{subsec:optical}
\begin{figure}
	\includegraphics[width=\columnwidth]{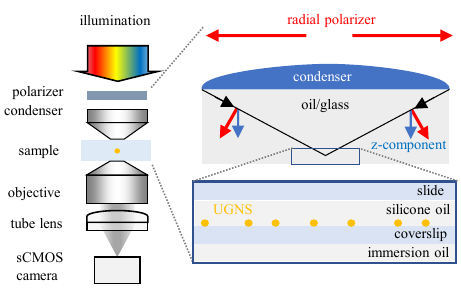}
	\caption{Sketch of the microscope set-up and investigated sample. Under Köhler illumination of selectable wavelength range, the sample is imaged onto a camera, using a 1.34\,NA condenser and a 1.45\,NA objective. In the illumination a rotatable linear polariser or a radial polariser are used, the latter generating a significant axially-polarised illumination component in the focal plane. The sample consists of individual gold nanospheres deposited onto a glass coverslip, surrounded by silicone oil index-matched to glass.}
	\label{fig:setup}
\end{figure}

We performed wide-field optical transmission measurements on a Nikon Ti-U inverted microscope, with illumination provided by a 100\,W tungsten halogen lamp, using bandpass filters of 40\,nm width and center wavelengths $\Lambda\in{\cone,\,\ctwo,\,\cthr,\,\cfour,\,\cfive,\,\csix}$. A 1.34 \NA\ oil-immersion condenser (Nikon MEL41410) focused the illumination to the sample. The transmitted light was collected by a 1.45\,\NA\ 100$\times$ oil immersion objective (Nikon MRD01905) and imaged by a 1.0$\times$ tube-lens onto the camera with a magnification of $M=100$. A motorised linear polariser in the collimated light path towards the condenser controlled the excitation polarisation angle \gamp, and a motorised slider switched between linear and radial polariser.\cite{AlabdullahAPL24} A scientific-CMOS (sCMOS) camera (PCO Edge 5.5) at the eyepiece port was used for image acquisition, having $2560\times2160$ pixels with a full-well capacity of $\Nfw=30000$ electrons. Illumination intensities and exposure times were chosen to result in pixel values around 50000 counts of the  16-bit digitizer range, which for the gain of 0.54 electrons/count corresponds to about 27000 electrons, close to \Nfw\ without entering a non-linear response range. For each wavelength, taken in ascending order, the sample was focussed, and then data were taken for linear polarisation at \gamp\ of  0, 45, 90 and 135 degrees, and radial polarisation. A sketch of the microscope set-up and sample geometry is shown in \Fig{fig:setup}, and more details are given in \Onlinecite{AlabdullahAPL24}. 

\subsection{Samples}\label{sec:sampleprep}

`Ultra-uniform' gold NPs of nominal spherical shape and mean diameter, $D=30\,\nm$, were obtained from NanoComposix. To observe small gold NPs with extinction microscopy, glass slides and coverslips (Menzel-Gl{\"a}ser, \#1.5) have to be optically clear, with no contamination or debris. This was achieved using a cleaning process involving sonication steps in toluene, acetone, a water rinse followed by treatment with boiling deionised (DI) water, and then storage in 30\% hydrogen peroxide for at least one day before preparation. A wet casting method was used for NP deposition, to prevent surface tension-induced aggregation during the sedimentation of NPs on the coverslips in a dry environment. The concentration of NPs in the suspension applied onto the glass surface was adjusted to achieve a surface density with individual NPs separated by a few microns. The stock concentration provided by the manufacturer was diluted accordingly. Slides and coverslips were rinsed with DI water, dried in a nitrogen flow, and 50\,\textmu L of the diluted NP suspension was evenly spread across the whole coverslip surface. The sample was placed in a 100\% humidity environment for one hour, to allow NP sedimentation via wet casting. Afterwards, the coverslip was gently rinsed with DI water, positioned vertically to allow excess water to drain, and dried in a nitrogen flow. Then, 25\,\textmu L of silicone oil (AP150, Wacker), index-matched to glass, was pipetted onto the coverslip to create an index-matched medium surrounding the NPs. A glass slide was placed over the oil-covered surface, and the coverslip was pressed manually to reduce the oil layer thickness. Excess oil at the edges of the coverslip was removed and the sample was sealed by applying nail varnish to the edges of the coverslip.

\subsection{Optical Extinction Measurements}\label{sec:sigmeas}

\begin{figure*}
\centering
\includegraphics[width=\textwidth]{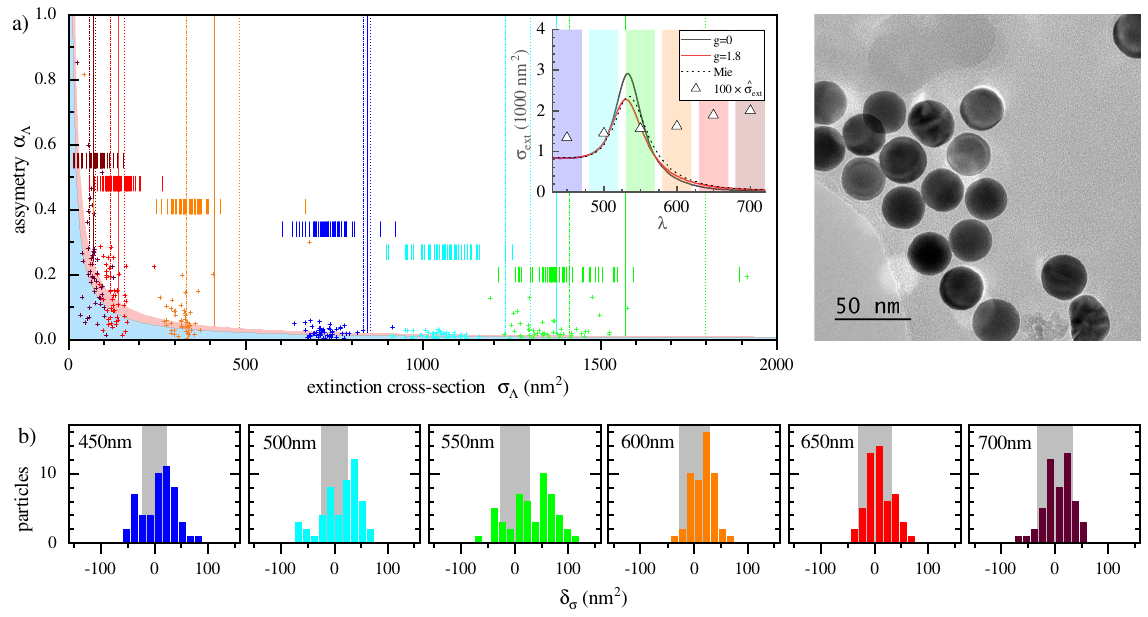}
\caption{Measured properties of $N=51$ UGNSs of $D=30\,\nm$ nominal diameter. a) Asymmetry \alam\ versus cross-section \siexl\ for $\Lambda=\{\cone,\,\ctwo,\,\cthr,\,\cfour,\,\cfive,\,\csix\}$, color-coded to reflect the center wavelength. Coloured crosses indicate the fitted parameters for linear polarisations using \Eq{eq:sinfit}. Coloured vertical bars indicate the measured \sirad\ for radial polarisation, at vertical positions chosen for clarity. A representative TEM image\cite{PayneNS20, PayneJCP21} of NPs from the same batch is shown on the right. The noise in \alam\ is shown as shaded areas for $\Lambda=\cone$ (bluish) and $\Lambda=\csix$ (reddish), estimated as $(\siexsb/\sqrt{1.5})/\siexl$, corresponding to the lowest and highest noise versus $\Lambda$, respectively. See \Sec{sec:optnoise} for further details about the noise in the fit parameters. Solid vertical lines indicate the calculated \siexl\ for a spherical gold NP for each  $\Lambda$, taking into account averaging over filter ranges, for the sample-averaged TEM diameter \DV, $27.94\,\nm$, and the OL permittivity dataset with $g=1.8$. The corresponding spectrum is given in the inset, showing additionally the case $g=0$. Dotted vertical lines indicate the same as solid, but calculated using Mie theory. Dash-dotted lines indicate the same as dotted, but for MP permittivity. The corresponding spectrum is shown by the dotted curve in the inset. Vertical colour bands indicate the filter ranges used in the experiment. The measurement error \siexsb, polarisation-averaged and FOV-averaged, are shown for each $\Lambda$, scaled by a factor $100$. 
b) Histograms of \delsig, for each $\Lambda$ as labelled, with gray bands showing $\pm\sqrt{2}\,\siexsb$, where the $\sqrt{2}$ factor accounts for the additional noise arising from the subtraction in \delsig.}
\label{fig:sigma_plots}
\end{figure*}

In \Onlinecite{PaynePRAP18} we detailed our procedure to quantitatively measure \siex, with additional information given in \Onlinecite{PaynePhD16}. In short, two \textit{brightfield} images called $I_1$ and $I_2$ are captured, which differ only by a lateral shift of the sample, typically 1 to 2\,\textmu m (i.e. a few optical resolutions). $I_1$ and $I_2$ are averaged in real time over a number, \Ni, of individual acquisitions to reduce shot noise. We compute $\Delta_{1,2}=1-I_{1,2}/I_{2,1}$ to yield two extinction images (with inverted contrast). To further reduce shot noise as well as systematic noise due to sensor electronic drift, we can average the extinction over a number of repetitions \Nr. For a NP centered in the areas $A_{1,2}$ of radius \ri\ in $I_{1,2}$, we determine its extinction as $2\siex=\int_{A_1}\Delta_1\mathrm{d}A+\int_{A_2}\Delta_2\mathrm{d}A$.
We calculate the shot-noise-limited standard deviation in the measured \siex\ as \cite{PaynePRAP18}
\be\label{eq:noise}
\siexsb=\frac{r_i\dpix}{M}\sqrt{\frac{\pi}{\Na\Nfw}}\,,
\ee
where $M$ is the magnification from the sample to the detector, $\Na=\Ni \Nr$ is the number of acquired frames, and $\dpix=6.5\,$\textmu m is the pixel pitch of the sensor. We previously used\cite{PayneAPL13} a measurement radius of $\ri=3\lambda/(2\NA)$ to analyse \siex, and the effect of \ri\ on \siex\ is characterised in detail in \Onlinecite{PayneSPIE19}. In the present work, we employ three values of \ri, specifically $\rs = 3\lambda/(2\NA)$, $\rh = 1.2\,\rs$, and $\rl = 0.8\,\rs$, with $\Ni=128$ and $\Nr=20$. At $\Lambda=550\,$nm, \Eq{eq:noise} yields $\siexsb\approx7.5\,\nm^2$ using \rs, while the noise seen in the extinction data at this wavelength is $15.6\,\nm^2$. The additional noise is attributed to surface roughness, debris, and/or residual sensor fluctuations, as analysed in detail in \Onlinecite{PaynePRAP18}. We note that measurements are taken over several different fields of view (FOV) and aggregated for the final statistics. The \siexsb\ reported are the polarisation-averaged values of the noise, averaged over the measured values from all the FOVs. We found that \siexsb\ varies by $\pm5\%$ across the illumination polarisations, and by $\pm 15\%$ across FOVs. We measure \siex\ for multiple wavelengths $\Lambda$, both for the linear polariser at multiple angles \gamp, denoted as $\siexl(\gamp)$, and for the radial polariser, denoted as \sirad. The noise \siexsb\ refers, in general, to the value specific to the data of each $\Lambda$, polarisation, and field-of-view (see \Sec{subsec:optical} for  $\Lambda$ and \gamp\ values used). The wavelength dependence of the noise is shown in the inset of \Fig{fig:sigma_plots}a.
 
We model the polarisation dependence of \siex\ for linear polarisation as
\be\label{eq:sinfit}
\sigma(\gamp)=\sigma\left\{1+\alpha \cos[2 (\gamp-\gamma)]\right\}\,,
\ee
where $\sigma$ is the average cross-section, and the polarisation dependence is given by a relative amplitude $\alpha\in[0,1]$, which encodes the dipolar in-plane asymmetry of the NP, with $\alpha=0$ indicating the absence of such asymmetry. The maximum of $\sigma(\gamp)$ is at the angle $\gamma\in[0,\pi]$, providing the orientation of the asymmetry in the sample plane.

The finite bandwidth of the filters used means the measured \siexl\ are averaged over the respective wavelength ranges. Additionally, while the transmission of the filters is approximately flat over their ranges, the illumination intensity and camera sensitivity are not. To take these features into account in our model of \siex, we compute weighted averages over each $\Lambda$ for each polariser angle, as
\be\label{eq:avgsigma}
\siexl = \frac{\int_\Lambda \,\wl)\,[\sia(\lambda)+\sisc(\lambda)] \,\mathrm{d}\lambda}{\int_\Lambda \wl\,\mathrm{d}\lambda}\,,
\ee
where $\wl = \il\cl$ is the weighting incorporating the normalized illumination spectral intensity, \il, and camera sensitivity, \cl, as functions of wavelength $\lambda$. The lamp spectrum is modeled as a blackbody radiator at temperature, $T = 2800\,$K, as described in Section 2.1.4 of \Onlinecite{PaynePhD16}. The camera sensitivity is known from the manufacturer and was digitized for this work. The wavelength dependence of \wl\ is shown in \Fig{fig:weights}. 

\section{Results and Discussion}\label{sec:Results}

This work demonstrates the enhanced capabilities of the new ON including radial polarisation in determining individual NP morphometries in 3D. We maintain the naming and definitions of variables and quantities used in our previous publications,\cite{PayneNS20,PayneJCP21} and introduce them in the text where needed. All analysis assumes that the refractive index of the NP environment in the samples is $\nmed=1.52$. The effect of variation of this index is discussed in \Sec{sec:fitindex}.

\subsection{Statistical Distributions of \siexl\ and \alam}
\label{sec:stats}
We first introduce relevant quantities used in the statistical description of the optical measurements. For each NP, \Eq{eq:sinfit} is fitted to the $\siexl(\gamp)$  measured for linear polarisation, determining its parameters $\sigma$, $\alpha$, and $\gamma$. Their standard errors $\hat{\sigma}$, $\hat{\alpha}$, and $\hat{\gamma}$ are determined using a Monte Carlo simulation where we add random Gaussian noise to the measured data, matching the experimentally determined noise, as detailed in \Onlinecite{PayneNS20} and \Sec{sec:optnoise}.

Given a set of $N$ NPs, numbered by the index, $i$, we identify the fit parameters using the subscript $\Lambda,i$, and define the sample-wide mean of $\sigma_{\Lambda,\,i}$ to be $\siexml = \frac{1}{N}\sum_{i=1}^N \sigma_{\Lambda,\,i}$, and the associated standard deviation  $\siexsl$. Equivalent quantities are defined for the other two parameters. Correcting for the measurement noise, the standard deviation $\siexsrl=\sqrt{\siexsl^2-\frac{1}{N}\sum_{i=1}^N \hat{\sigma}_{\mathrm{\Lambda},i}^2}$ represents the distribution of NP cross-sections originating from their sizes and shapes only. For the relative amplitude parameter, the measurement error $\hat{\alpha}_{\Lambda,i}$ varies strongly depending on the parameter values, so that we refrain from such a correction and directly use $\ais\Lambda = \aisl$.

The UGNS are highly uniform in size and shape as previously measured by HAADF--STEM.\cite{PayneJCP21} Specifically, we considered the projected diameter \DP\ and the PAR using the shorter, $b'$, to longer, $a'$, semi-axes observed in plan-view STEM, and found distributions characterised by mean and standard deviation values of $(\DPb\,\pm\,\DPh) = (28.04\,\pm\,1.54)$\,nm for the diameter and $0.977\,\pm\,0.024$ for the PAR. 
The median $M\{.\}$ over the $N$ NPs analyzed is found to be $M\{\DP\} = 27.75\,\nm$ and $M\{\mathrm{PAR}\}=0.985$. From this prior knowledge, we can expect for our optical measurements: (1) a narrow distribution of both \siexl\ and \alam, (2) a small in-plane asymmetry $\aiml \ll 1$ for all $\Lambda$ where $\siexml\gg\siexsb$, (3) \siexml\ to be close to the theoretically predicted value for a GNS of the expected size.

\Fig{fig:sigma_plots} presents \siexl\ and \alam\ for measurements of the $N=51$ nominally $D=30\,$nm UGNS for all $\Lambda$ of the experiment. To mitigate small differences in focus between the different fields of view, \siex\ was determined using the larger radius \rh\ (see \Sec{sec:sigcomp} for more details). One must correct \siex\ obtained with a finite \ri\ to compensate for the signal outside of the integration area. The correction factor to \siex\ was determined to be $c_\mathrm{l} = 1.12$, $c_\mathrm{s} = 1.06$, and $c_\mathrm{h} = 1.02$,  for \rl, \rs, and \rh, respectively. In \Fig{fig:sigma_plots} we show \siex\ as measured and the modelled extinction divided by $c_\mathrm{h}$. Narrow distributions of \siexl\ are observed for all $\Lambda$. Small in-plane asymmetries $\alam<0.2$ are found for about 95\% of the NPs at $\Lambda = \lfour$ and below, and for about 80\% of NPs at $\Lambda = \lfive$. At $\Lambda=\lsix$ instead, \siex\ is only a few times larger than the noise, which results in a wide distribution of \alam\ estimated by the shaded areas. Hence, the sample exhibits a low optical asymmetry, with some 10\% outliers.

Looking at the simulations in the inset of \Fig{fig:sigma_plots}, we note that $\Lambda=\lthr$ covers the higher wavelength side of the LSPR of a spherical gold NP, suggesting that accurately modelling the width of the channels will be relevant for the morphology retrieval. With increasing NP asymmetry, the LSPR polarised along the longer axis (the longitudinal mode) red-shifts, increasing its overlap with the $\lthr$ channel, and thus providing a sensitive probe of asymmetries. We find $\aim{\lthr} = 0.037\,\pm\,0.039$, with some 80\% of NPs showing $\alpha_{\lthr}<0.054$, and the rest clearly asymmetric with $\alpha_{\lthr}$ up to 0.194. For asymmetric NPs, the LSPR will move into the increasingly red channels (see simulations in \Onlinecite{PayneNS20} Fig.\,2), increasing their \siexl\ and \alam. Looking at the exemplary TEM image, some 10-20\% of significantly non-spherical NPs can be expected. Their red-shift can be significant, contributing to the values of $\alpha_{\lfive}$ and  $\alpha_{\lsix}$ well above the noise. We also note that the $\Lambda=\{\lone,\,\ltwo\}$ channels are rather insensitive to asymmetry, due to the dominant absorptive part of the permittivity, $\Im\{\ep\}\approx 5$, with $\Re\{\ep\}$ similar to \emed, suppressing shape-dependent effects, so that  $\sigma_{\lone}$ and $\sigma_{\ltwo}$  are good reporters of the NP volume.

Let us now discuss \siex\ measured using the radial polariser, \sirad, see vertical bars in \Fig{fig:sigma_plots}, which we compare to the fitted \siexl\ for linear polarisation shown as crosses. We recall that the electric field at the sample has both in-plane and out-of-plane components, and we consider\cite{ZilliACSP19, AlabdullahAPL24} the relative strengths of the components along the $x$, $y$, and $z$ axes, respectively as $E=[E_\mathrm{x},\, E_\mathrm{y},\, E_\mathrm{z}]$. The intensity in the back focal plane (BFP) of the condenser varies spatially, with a maximum at the center, determined by the lamp, diffuser and lenses in the illumination path, as characterised in our previous work.\cite{ZilliACSP19} Using the known spatial dependence, we can calculate the intensity components at the sample polarised along the $x$, $y$, and $z$ axes, normalized to have unity sum. For a linear polariser along $x$ in the condenser BFP, we find $I_\mathrm{lin} = [0.826,\,0.007,\,0.167]$, corresponding to field components $E_\mathrm{lin} = [0.909,\,0.084,\,0.409]$ given by the square root of the intensity components.  For radially polarised input to the BFP, we have $I_\mathrm{rad} = [0.333,0.333,0.334]$ corresponding to $E_\mathrm{rad} = [0.577,\,0.577,\,0.578]$, showing that the $z$ component is increased in the radially polarised case.  

For a spherical NP, the polarisation direction should not affect the measured \siex, and thus we expect the difference $\delta_\sigma = \sirad-\siexl$ to be negligible. The histograms of \delsig\ are shown in \Fig{fig:sigma_plots}b, and we find for channels with large cross-sections $\Lambda=\{\lone,\,\ltwo,\,\lthr\}$ a shift towards $\delsig>0$ of relative magnitude $\delsig/\siexl$ below 3\%, while for the other channels \delsig\ is dominated by noise, shown as gray bars. To explain this small shift, we recall the ``long shadow'' effect,\cite{PayneSPIE19} which arises from the illumination covering an oblique angular range while referencing to the intensity transmitted through the imaged sample plane. The \NA\ of the condenser and the intensity distribution in its BFP provide the weighting of the illumination angles at the sample, from which we calculate the long-shadow factor used. However, the transmission of the condenser and objective for oblique angles is polarisation-dependent, due to the reflections on the lens surfaces, which are generally lower for p polarisation compared to s polarisation. Radial polarisation is purely p polarised, while linear polarisation contains equal parts of s and p polarised light. Considering that the long-shadow factor is increasing with incidence angle,\cite{PayneSPIE19} the higher transmission at higher angles for radial polarisation creates an illumination with a higher long-shadow factor. This systematic error can be approximately compensated by reducing the measured \sirad\ by a factor $0.97$, having an effect explored below. One could also speculate that a preferential orientation of the NPs on the surface creates the difference. However, we would expect in such a case that NPs preferentially lie flat on the surface, providing a larger binding surface, which would lead to a higher cross-section for in-plane fields, opposite to what is observed.
\begin{figure}
	\includegraphics[width=\columnwidth]{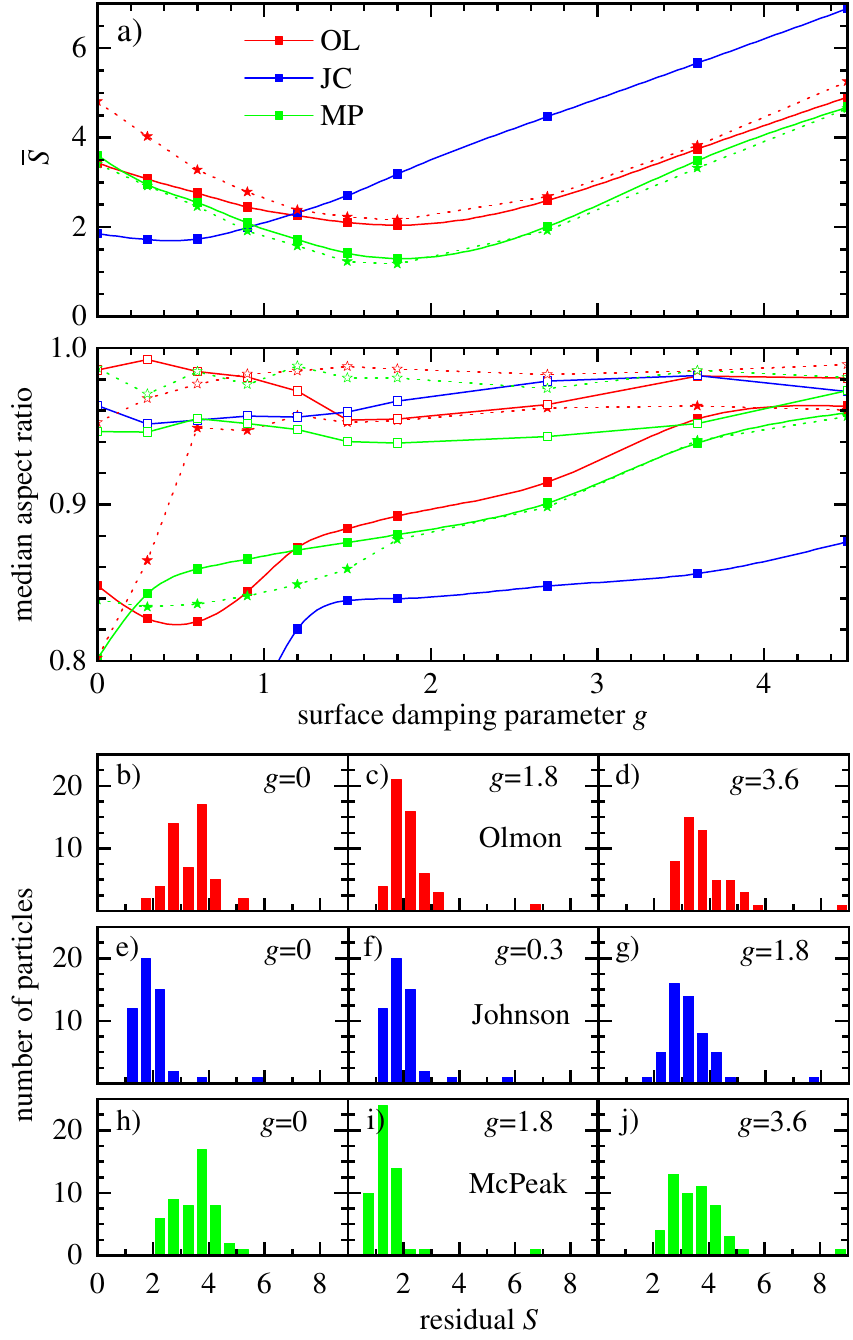}
	\caption{ \Sbar\ and selected statistics of $S$ for $N=51$ UGNS of nominal $D=30\,$nm for varying surface damping parameter $g$ and different $\epsilon$ datasets as indicated by color. (a) Top: \Sbar\ versus $g$. Solid lines and squares for $\ep(\lambda)$. Dotted lines and stars for $\ep^s(\lambda)$ with $0.97\times$ correction for $\siex(\gamr)$ data. Bottom: median of the fitted average aspect ratio, for MAR (filled symbols) and PAR (empty symbols). Solid and dotted line designations same as Top. (b--j) Histograms of $S$, with (b--d) using the OL dataset with $g=\{0,\, 1.8,\, 3.6\}$, (e--g) JC dataset with $g=\{0,\, 0.3,\, 1.8\}$, (h--j) MP dataset with $g=\{0,\, 1.8,\, 3.6\}$. The values of $g$ shown are at $g=0$, the minimum of $\Sbar(g)$ and a larger $g$.}
	\label{fig:ErrorOverview}
\end{figure}

\subsection{Morphometric analysis}\label{sec:morph}
We introduced the morphometric analysis method, i.e. solving the inverse problem, in \Onlinecite{PayneNS20}. We summarize the key points here. \sisc\ and \sia\ are discretely calculated over a multidimensional space with parameters ({\brat, \crat, $\phi$, $\theta$, $\psi$}), where the illumination intensity spectra of the color channels $\Lambda$ and the selected permittivity are taken into account. We then generate interpolants for \sisc\ and \sia\ using the calculated grid. We compare the measured values of a given particle $i$ to the calculated values for each point on the grid using the normalized error,
\be\label{eq:err}
S^2=\frac{1}{n} \sum_{\gamp,\,\Lambda} \left(\frac{\siexl(\gamp)-\sial(\gamp)-\eta\siscl(\gamp)}{\siexsbl(\gamp)}\right)^2\,,
\ee
where we dropped the index $i$ for brevity. $n$ is the number of measurements in the sum, which includes the radially polarised measurement $\gamp=\gamr$, and $\siexsbl(\gamp)$ is the measurement noise associated with \siexl(\gamp). The factor $\eta$ is the fraction of scattered light which is not collected by the objective and thus contributes to the measured extinction. For our experimental configuration we calculated \cite{PaynePhD16} $\eta=0.646$.

Since the right side of \Eq{eq:err} is a fourth-order polynomial in $V$ we can minimise $S^2$ with respect to $V$ analytically to obtain $S$ and $V$ for the specific dataset and obtain the volume-equivalent diameter, $\DV=\sqrt[3]{6V/\pi}$. Points on the grid where $S^2$ is smaller than a suitable cut-off are then used as initial guesses for a gradient descent employing the \sisc\ and \sia\ interpolants, each providing a solution. We use the solution with the minimum $S^2$ as descriptor of the morphometry and orientation of the NP. To constrain retrieved morphometries, we have previously applied additional penalties using prior knowledge, e.g. ensemble specifications from the NP manufacturer. We do not implement this here to avoid the influence of systematic errors in such specifications.

The six spectral channels used improves the interrogation of the LSPR spectral shape compared to \Onlinecite{PayneNS20}. Hence, we re-examine here the best choice of the permittivity \ep, defined by the minimum of
\be\label{eq:S}
\Sbar=\sqrt{M\{S^2\}}\,.
\ee
Since $S$ is normalized by $\siexsb(\gamp)$ in \Eq{eq:err}, we would expect $S=1$ for the correct model, without taking into account the number of fit parameters and the number of data points to be fitted per NP. Considering that we use $p=6$ free parameters to fit the $n=30$ measurements (four linear and one radial polarisation at 6 wavelengths), this expectation is reduced by the number of degrees of freedom $n-p$ to $S\approx\sqrt{(n-p)/n}\approx 0.9$.

\begin{figure*}
	\centering
	\includegraphics[width=\textwidth]{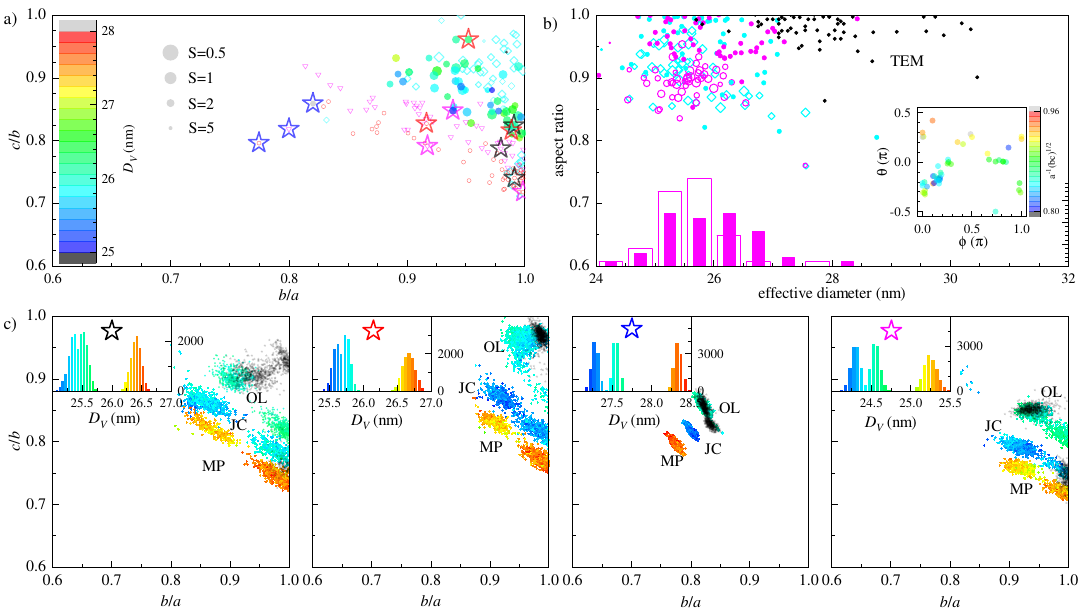}
	\caption{Morphometric results for $N=51$ nominally $D=30\,$nm UGNSs. (a) Retrieved ARs for OL $g=1.8$ are shown as filled symbols with a colour indicating the fitted diameter \DV. ARs retrieved using only $\siexl(\gamp)$ are shown as cyan empty diamonds. Also shown are retrieved ARs for JC $g=0.3$, (magenta empty triangles), and for MP $g=1.8$ (red empty circles). (b) MAR versus \DV\ for OL $g=1.8$ (magenta empty circles), with the histogram showing the number distribution of \DV\ (empty) and \DP\ (filled). Corresponding PAR are shown as filled symbols versus \DP. Results using only \siexl(\gamp) are shown as diamonds. The black filled squares indicate TEM measurements PAR versus \DP. Inset: pitch ($\theta$) and roll ($\phi$) angles showing the particle orientation for OL, with a symbol colour encoding the MAR.  The size of the circles in (a) and the empty symbols in (b) is given by $S^-=1/(1+S)$, as indicated in (a), with larger size indicating lower fit error $S$. (c) Simulated effect of the measurement noise on the retrieved ARs for individual NPs indicated by the colored stars in (a). Results for 1000 realisations of Gaussian noise with \siexsb\ standard deviation added to the measured extinction data are shown. Inset : \DV\ histograms, providing a colour scale for the symbols. Results for OL, JC, and MP at $g$ of $1.8$, $0.3$, and $1.8$, respectively, with corresponding results labelled. Black symbols are results using only $\siexl(\gamp)$ for OL.}
	\label{fig:30nmFitPlot}
\end{figure*}

\Sbar\ is shown for the data analysis as discussed to this point by the solid curves in \Fig{fig:ErrorOverview}a as a function of the surface damping parameter $g$ for each of the experimental $\epsilon$ datasets. We find minimum values of $\Sbar=1.29$ at $g=1.8$ for the MP dataset, $\Sbar=2.05$ at $g=1.8$ for the OL dataset, and $\Sbar=1.72$ at  $g=0.3$ for the JC dataset. Compared to our results in \Onlinecite{PayneJCP21}, \Sbar\ is more sensitive to $g$, due to the lower noise and larger number of measurements used. The histograms of $S$ are shown in \Fig{fig:ErrorOverview} for each dataset and $g$ at the minimum of $\Sbar(g)$ and either side of it. At the minimum, a nearly Gaussian shape is observed. The distributions of $S$ for each dataset display a few outliers of large $S$, indicating some NPs are not well described by the model.

Notably, the smallest \Sbar\ is found for MP, which indicates the best choice since it represents the model which best fits the data. On the other hand, we see in \Fig{fig:ErrorOverview}, \ref{fig:Sbar_SM}, and \ref{fig:Ratios_SM}, that OL retrieves the roundest shapes as shown by $M\{\mathrm{MAR}\}$, closest to the shape measured in TEM. Excluding the radial polarisation data yields the lowest \Sbar, for all permittivities (see \Fig{fig:Sbar_SM}). This can be rationalised noting that the radial polarisation probes the shape in the out-of-plane direction, restraining the fit from using this shape to create spectral shifts while the in-plane shape is constrained by $\siexl(\gamp)$. 

In principle, all permittivity datasets should be consistent apart from surface damping, as they have all been measured on gold films of different crystallinity. While this is the case for OL and JC, with JC corresponding to OL with $g\approx1$, reflecting the lower crystallinity of the film measured in JC compared to the single crystal data in OL, MP corresponds to a higher plasma frequency, leading to a blueshift of the LSPR of a spherical NP by about\cite{PayneJCP21} 4\,nm. Using MP in the fit thus yields higher NP asymmetry, which compensates the blueshift by a redshift due to non-spherical shape. 

The results of the morphometric fits for the $N=51$ nominally $D=30\,\nm$ UGNS are shown in panels (a--b) of \Fig{fig:30nmFitPlot}. The retrieved NP shapes for OL $g=1.8$, shown as filled circles, are slightly non-spherical, with $\cbrat =  0.91 \pm 0.032$, $\crat = 0.82 \pm 0.12$, and $\brat = 0.96 \pm 0.04$, where the errors are standard deviations of the ensemble, and a mean (median) MAR of 0.89 (0.90). When excluding \sirad, the results show slightly rounder NP shapes, with $\cbrat = 0.88 \pm 0.07$, $\crat = 0.85 \pm 0.07$, $\brat = 0.97 \pm 0.03$, and a mean (median) MAR of 0.91 (0.91). Here, the shapes form two clusters, with the smaller cluster having \cbrat\ lower than any shapes retrieved when including \sirad. We find $\DV = (25.6\pm0.5)\,$nm  (median 25.6\,nm) over the ensemble when including \sirad, showing a systematic deviation of about $2$\,nm, or 8\%, to the TEM results. The results for the JC and MP datasets generally show retrieved shapes which are more widely distributed and of stronger non-sphericity compared to OL, and MP retrieves nearly $1\,\nm$ larger average diameters. We explore in \Sec{sec:fitcomp} the effect of the different corrections and selections introduced on the shape retrieval in more detail.

Notably, the projected shapes are generally rounder and of larger diameter, see filled circles (PAR) and filled histogram (\DP) in \Fig{fig:30nmFitPlot}b. This can be interpreted in two ways: (i) since the in-plane asymmetry is measured with higher precision via the \gamp\ dependence, the PAR has less error and thus reflects the correct aspect ratio in all three dimensions of the NP, (ii) the NPs are orientated on the surface in a way to appear more round in-plane. Considering the nearly spherical shapes, a physical mechanism which would result in (ii) both for TEM and optical measurements is not obvious.

In \Fig{fig:30nmFitPlot}c, we examine the uncertainty in the retrieved ARs and \DV\ due to measurement noise for four selected NPs, indicated by the colored stars in \Fig{fig:30nmFitPlot}a. The analysis was done for all three permittivity datasets with $g$ minimizing \Sbar. We simulate the effect of the measurement noise by adding a realisation of Gaussian random noise with a standard deviation of $\siexsbl(\gamp)$  (see inset of \Fig{fig:sigma_plots}a) to the measured $\siexl(\gamp)$, and then repeat the morphometric analysis. Each set of scatter data in each of the four panels represents 1000 noise realizations.  The ARs often exhibit  bimodal distributions, with the greatest separation between modes apparent for MP and JC, while the width of the individual modes is typically below $5\%$. Notably, the modes often are close to $b=a$ or $c=b$, allowing to achieve zero in-plane asymmetry (which is defined with high precision by the linear polarisation data) for appropriate NP orientations. Using synthetic \siex\ data generated using the Rayleigh-Gans model and the specified experimental corrections, such multimodal results are not observed,\cite{PayneNS20} suggesting that they are caused by systematic measurement errors. Notably, the sample distributions in \Fig{fig:30nmFitPlot}a for each permittivity mirrors the spread of the noise distributions for the rounder NPs in \Fig{fig:30nmFitPlot}c for each permittivity, which points to a homogeneity of the sample, as well as this systematic effect. We will return to this topic again later in the text. Generally, for near-spherical NPs, AR retrieval is less precise due to the quadratic parameter dependencies close to symmetry points of the the multi-dimensional parameter space.\cite{PayneNS20} Significantly asymmetric NPs instead yield more precise retrieval, as exemplified by the blue-starred NP. In this case, precision is better than $3\%$, corresponding to 0.84\,nm length along the NP axes. For further insights into this effect, see \Sec{sec:optnoise}. Variations in \nmed\ result in systematic changes in $c/b$ of about twice the index change, as discussed in \Sec{sec:fitindex}. For a realistic refractive index accuracy of 0.01, the corresponding $c/b$ systematic error is 2\%. Remarkably, the retrieved \DV\ shows a precision of $(0.04 - 0.1)\,\nm$ standard deviation for the selected NPs.

The systematic deviation of about $2\,\nm$ between the mean of the retrieved \DV\ and that found by TEM might be due to several factors. A known systematic error is the quasi-static approximation used in the model. Considering the wavelength in vacuum of the LSPR for a spherical gold NP of about $\lambda_0 = 530\,\nm$, the wavelength in the medium of $\nmed=1.52$ is $\lambda_0/\nmed \approx 350\,\nm$, which is about 12 times the nominal diameter of the NPs. This suggests that taking into account retardation, which leads to radiative damping and red-shift of the LSPR, might be relevant to achieve accuracies below 10\% for the investigated NPs. Indeed, the extinction spectrum of a gold NP with the TEM-measured average diameter $\DV = 27.94\,$nm, calculated including retardation using Mie theory (see inset of \Fig{fig:sigma_plots}a, dotted black line), shows an about 4\% increase and 5\,nm red-shift of the LSPR peak, see also \Sec{sec:fitmodel}. The vertical lines in \Fig{fig:sigma_plots}a show the corresponding weighted cross-sections (solid for Rayleigh-Gans, dotted for Mie) for all colour channels. 
We would expect that the redshift would result in rounder retrieved NPs, while the increase would result in an about 1\% smaller \DV\ considering the approximate scaling of the cross-section with the volume. 

\begin{figure*}
	\centering
	\includegraphics[width=\textwidth]{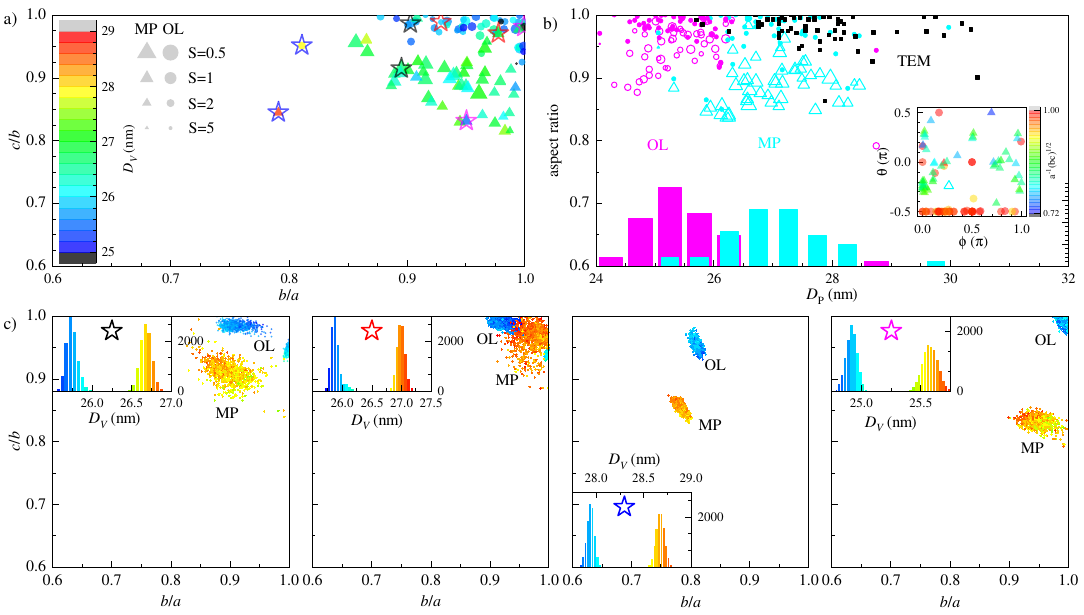}
	\caption{Similar to \Fig{fig:30nmFitPlot}, but for OL and MP using $\ep^s(\lambda)=\ep(\lambda+\delta\lambda)$ with $\delta\lambda=5\,\nm$ and the $0.97\times$ correction for $\siex(\gamr)$.
		(a) Retrieved ARs shown as filled symbols (MP triangles, OL circles) with a color indicating the fitted diameter \DV. (b) MAR (empty symbols) and PAR (filled symbols) versus \DP, with the histogram showing the number distribution of \DP, with OL shown as magenta circles\,/\,bars and MP as cyan triangles\,/\,bars. The black filled squares indicate TEM measurements PAR versus \DP. Inset: pitch ($\theta$) and roll ($\phi$) angles showing the particle orientation.  (c) Simulated effect of the measurement noise on the retrieved ARs for individual NPs indicated by the colored stars in (a). Results for 1000 realisations of Gaussian noise with \siexsb\ standard deviation added to the measured extinction data are shown. Inset: \DV\ histograms, providing a colour scale for the symbols. 
	}
	\label{fig:FitPlotShift}
\end{figure*}

To investigate this further while keeping the Rayleigh-Gans model, we approximate the effects of retardation by spectrally shifting the permittivity with respect to $\lambda$ to obtain $\ep^s (\lambda)= \ep(\lambda+\delta\lambda)$, with $\delta\lambda = -5\,\nm$, and repeating the ON analysis using $\ep^s$ and the $0.97\times$ correction for $\siex(\gamr)$. The resulting dependence of \Sbar\ and the median MAR on $g$ is given as dotted lines in \Fig{fig:ErrorOverview}a for OL and MP, showing that the $g$ at which \Sbar\ reaches a minimum is rather unaffected, while the median MAR increases. The corresponding ON results are given in \Fig{fig:FitPlotShift} for OL and MP. The retrieved shapes are rounder for both datasets, with OL showing clearly round shape retrieval with median MAR of $0.954$, compared to $0.892$ for MP. Meanwhile, the median diameter for MP is $M\{\DV\}=26.72\,\nm$, larger than $25.77\,\nm$ for OL. For OL, nearly all of the asymmetry is in \brat\, corresponding to a prolate particle, and most orientations show $\theta\approx-\pi/2$, corresponding to a prolate NP standing on end, i.e. its long axis is normal to the surface. This suggests a systematic effect, where a negligible in-plane polarisation-dependence combines with a constrain by \siex(\gamr). MP by comparison shows a less constrained distribution of orientations, with widely distributed tilts $\theta$.

The projected aspect ratios PAR versus the corresponding diameters \DP\ are also shown in \Fig{fig:FitPlotShift}b. The median PAR is $M\{\mathrm{PAR}\}=0.987$ for OL and 0.981 for MP, with its dependence on $g$ given in \Fig{fig:ErrorOverview}a. These values are significantly higher than those of MAR. The histograms of \DP\ are shown and the median for MP is $M\{\DP\}=27.02\,\nm$, larger than $25.37,\nm$ for OL. Remarkably, both PAR and \DP\ for MP are in good agreement with the TEM results ($M\{\mathrm{PAR}\}=0.985$, $M\{\DP\}=27.75$\,nm).

Furthermore, for the two rounder particles in \Fig{fig:FitPlotShift}c, OL shows distinctly bimodal distributions , while MP returns unimodal distributions. For these examples, at one standard deviation from the mean, the precision in \DV\ is improved to $(0.04 - 0.07)\,\nm$, and the precision in AR is (0.37\% - 2.8\%) for OL and (0.45\%-2\%) for MP, depending on the NP.
\section{\label{sec:Conc} Conclusions}
We have presented an enhanced ON method with optical cross-sections measured using six spectral channels, and both linear and radial polarisation to probe both in-plane and out-of-plane NP response. For nominally $30\,\nm$ ultra-uniform gold nanospheres, we demonstrated significant improvement in the precision and accuracy of the method. Precision in the size and shape has been improved to below $0.1\,\nm$ and $5\%$, respectively. We examined the effect of the permittivity dataset on the morphometry retrieval and found that McPeak {\it et al}.\cite{McPeakACSP15} with surface damping factor $g=1.8$ yields the best results.
We find that the small effect of retardation for these NP, which are some 12 times smaller than the light wavelength in the surrounding medium, provides a LSPR red-shift of some 5\,nm. This is significant due to the high precision of the measurements, and correcting for it improves  accuracy to within 5\% in shape and size using TEM measurements as reference. We however note that this accuracy is contingent on the permittivity used to describe the material response.

Future development could see the use of numerical models, capable of treating particles outside of the limitations of analytical models. Another approximation limiting the accuracy is the use of a local permittivity to describe the nanoparticle optical response, which is known to break down on the sub-nm scale in general. Again, more complex theory models\cite{ChristensenPRL17} can be used to improve on this. The precision of the method is so high that small systematic effects in the experimental setup, such as the angular resolved transmission and the angular intensity distributions, are relevant to achieve an accuracy comparable to, or better than, the precision. These can be corrected by further characterisation of the instrumentation. 

Overall, this study shows the ability of the ON to complement TEM, and as a stand-alone methodology for accurate NP morphometric studies. It also offers an interesting approach to determine the permittivity of materials on the nanoscale, and to uncover non-local effects.

\begin{acknowledgments}
F.A. thanks Al-Furat Al-Awsat Technical University (ATU) for the financial support toward her Ph.D. studies. Parts of this work were funded by the UK Research Council EPSRC Impact Acceleration Account of Cardiff University. The microscope set-up was supported by the UK Research Council EPSRC (grants n. No. EP/I005072/1 and EP/M028313/1). We thank Iestyn Pope for technical support in the experiments and in hardware development of the radial polariser.
\end{acknowledgments}

\section*{Author contributions}

L.P., W.L. and P.B. conceived the work. F.K. prepared the samples and performed the optical measurements. L.P, P.B. and W.L. developed the numerical model and fitting methods. L.P. performed the numerical simulations and data fitting.  All authors contributed to the data interpretation and writing of the manuscript.

\appendix

\section{Illumination spectrum and camera sensitivity}\label{sec:weights}

\begin{figure}[h]
	\centering
	\includegraphics[width=\columnwidth]{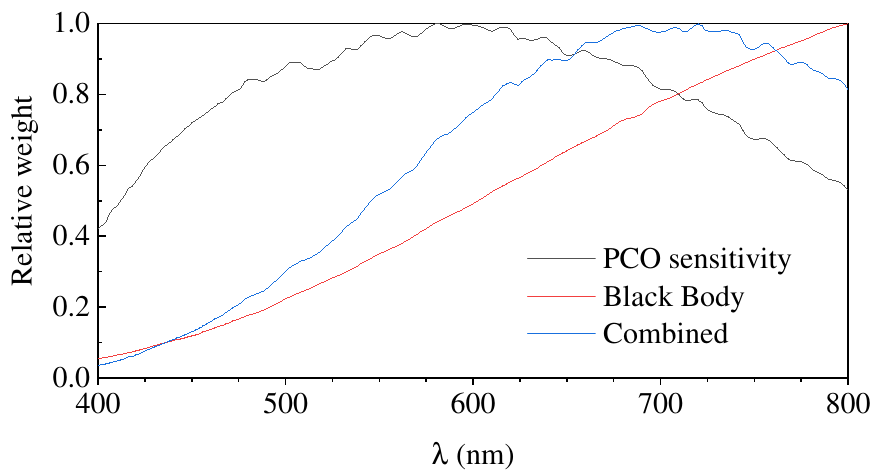}
	\caption{Plot of \il\ (red curve), \cl\ (black curve), and \wl\ (blue curve).}
	\label{fig:weights}
\end{figure}

The illumination setup in this experiment uses a halogen lamp as source, operated at close to max setting (100\,W), with 40\,nm bandwidth filters to provide coarse wavelength selection in the experiment. While the filters exhibit close to flat transmission across their bandwidth, the lamp spectrum is varying over the filter ranges, specifically in the green-blue region. To take this into account, we approximate the lamp spectrum as a blackbody radiator as in Section 2.1.4 of \Onlinecite{PaynePhD16}. One must also consider the spectral properties of the detection. The PCO Edge 5.5 scientific CMOS (sCMOS) exhibits a quantum efficiency (QE) depending on wavelength, which was characterised by the manufacturer and was digitised for this work. \Fig{fig:weights} shows the calculated blackbody lamp spectrum \il, the camera sensitivity \cl, and the combined dependence $\wl = \il \cl$, normalized for visualisation purposes.

\section{Projected ellipsoidal parameters}\label{sec:project}

Given the rotation angles ($\phi$, $\theta$, $\psi$) and semi-axis lengths ($a,b,c$) of the fitted ellipsoid, as introduced in \Sec{sec:perm}, we can calculate its projection along the $z$ axis in the laboratory frame, which is normal to the substrate surface, as follows.\cite{HartleyBook03} We introduce
the ellipsoid matrix
\be M=\begin{bmatrix}
	a^{-2} & 0 & 0\\
	0 & b^{-2} & 0\\
	0 & 0 & c^{-2}\\
\end{bmatrix} \ee
describing the ellipsoid surface as points $\mathbf{r}$ respecting $\mathbf{r}\cdot M \mathbf{r}=1$, and rotate it into the lab frame,
\be \hr M \hrt  = \begin{bmatrix}
	A & \mathbf{b} \\[4pt]
	\mathbf{b}^T & d \end{bmatrix}\ee
using the rotation $\hr=R_\psi R_\theta R_\phi$ as introduced in \Sec{sec:perm}. The rotated matrix is expressed by an in-plane matrix $A$, a vector $\mathbf{b}$, and a scalar $d$ as shown. We proceed by introducing the projected ellipse matrix
\be Q = A - \frac{\mathbf{b} \mathbf{b}^T}{d}\ee
which we diagonalise to determine its two eigenvalues $\lambda_1 < \lambda_2$, which define the in-plane semi-axis lengths $a' = 1/\sqrt{\lambda_1}$ and $b' = 1/\sqrt{\lambda_2}$.

\section{Noise of linear polarisation fit parameters}\label{sec:optnoise}

The extinction cross-sections, \siex, measured for different linear polariser angles, \gamp, are fitted with \Eq{eq:sinfit} to retrieve the in-plane dipolar asymmetry, $\alpha$, its orientation, $\gamma$, and the average cross-section, $\sigma$. 

To determine how the experimental noise, \siexsb\, arising from photon noise and sample background structure, affects the fit parameters, we simulate realisations of $\siex(\gamp)$ for $\Npol\in[3,4,6]$, asymmetries $\als\in[0,0.1,0.2,0.3,0.4,0.5,0.8,1.0]$, and orientations $\gams\in[0,10,15,20,30,45,90]$\,degrees. We generate $\siex(\gamp,\als,\gams)$ by adding Gaussian noise of standard deviation $\siexsb=0.1\sigma$ to the results of \Eq{eq:sinfit}, and fit the data using \Eq{eq:sinfit}, for $N=10000$ noise realisations. 

\begin{figure*}
	\centering
	\includegraphics[width=\textwidth]{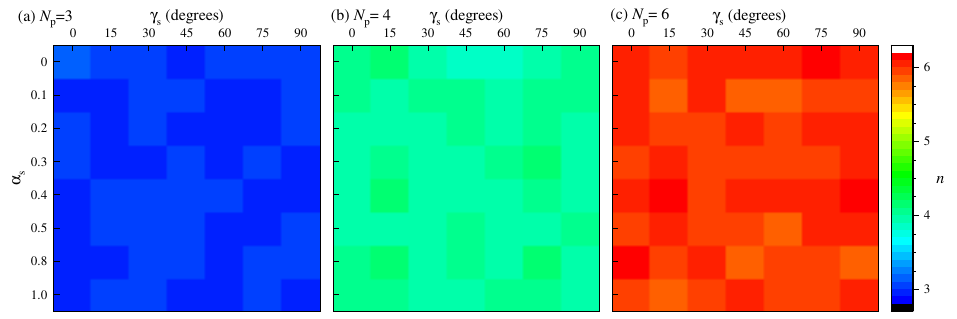}
	\caption{Effective number of measurements $n$ in the noise in $\sigma$, as function of $\als$ and \gams, determined from fits to simulated extinction data using \Eq{eq:sinfit} ($n = (\siexsb/\hat{\sigma})^2$, see text). Results are shown for 10\% relative noise in \siex, for different numbers $\Npol$ of equidistant polariser angles across $180^\circ$, with panels (a), (b), and (c) using $\Npol=3$, $4$, and $6$ angles, respectively.}
	\label{fig:signoise}
\end{figure*}

For small asymmetries \als, $\sigma$ is effectively the mean of $\siex(\gamp)$ over all $\gamp$, so that the noise in $\sigma$ should be given by $\siexsb/\sqrt{\Npol}$, where \Npol\ is the number linear polariser angles used. We accordingly define an effective number of measurements $n(\als,\gams) = (\siexsb/\hat{\sigma})^2$ for $\sigma$, and the simulation results shown in \Fig{fig:signoise} confirm that $n$ is close to $\Npol$, notably not only for $\als=0$, but across all \als.

\begin{figure*}
	\centering
	\includegraphics[width=\textwidth]{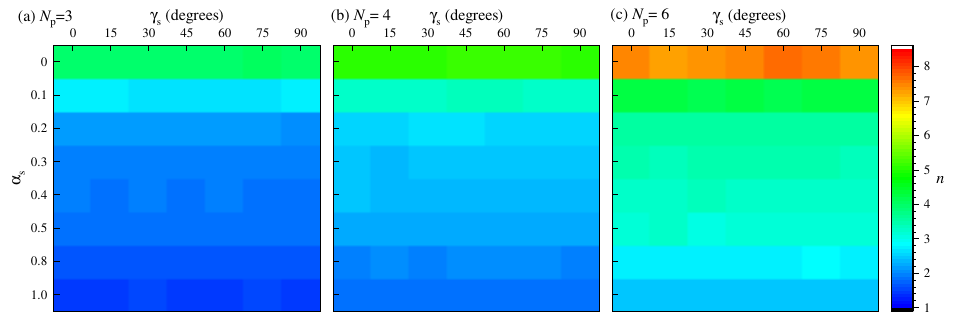}
	\caption{As \Fig{fig:signoise}, but for $\alpha$, using $n = (\siexsb/(\sigma \halpha))^2$.}
	\label{fig:optnoise}
\end{figure*}

For $\alpha$, the effective number of measurements is defined as $n(\als,\gams) = (\siexsb/(\sigma \halpha))^2$, noting that the oscillation amplitude is $\sigma\alpha$, and the simulation results are shown in \Fig{fig:optnoise}. Naively, one could expect $n = \Npol-2$, considering that we measure 2 additional parameters apart from $\alpha$. This is also approximately the case for $\als$ around 0.3-0.5. For small $\als<0.3$, $n$ increases, reaching about twice that value at $\als=0$. This is attributed to the choice $\alpha \leq 0$, controlling the sign of $\alpha$ by a 90 degree shift of the angle $\gamma$. Therefore, the distribution is narrowed, at the expense of noise in $\gamma$. For large $\alpha>0.5$ instead, $n$ decreases by about 1. We attribute this to a reduction of the noise in $\gamma$ at the expense of the noise in $\alpha$.

\begin{figure*}
	\centering
	\includegraphics[width=\textwidth]{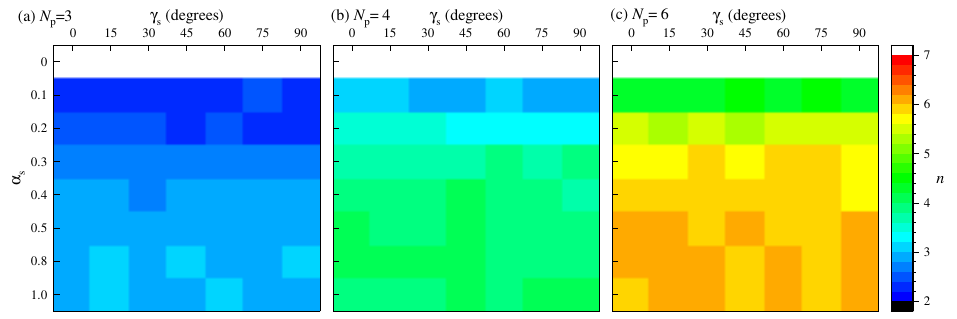}
	\caption{As \Fig{fig:optnoise}, but for $\gamma$, using $n = 0.5(\siexsb/(\sigma \alpha \hgamma))^2$, with $\gamma$ in radians, and \hgamma\ calculated as circular error.\cite{YamartinoJCAM84}}
	\label{fig:gamnoise}
\end{figure*}

For $\gamma$, the noise $\hgamma$ is calculated in radians using a circular statistics \cite{YamartinoJCAM84} with the periodicity of 180 degrees, and the  effective number of measurements is defined as $n(\als,\gams) = 0.5(\siexsb/(\sigma \alpha \hgamma))^2$, noting that the oscillation amplitude is given by $\sigma \alpha$. The simulation results shown in \Fig{fig:gamnoise} indicate that $n$ is about \Npol, except for small $\alpha$, where noise in $\gamma$ becomes so large that $\hgamma$ is limited by the periodicity.

Interestingly there is no clear dependence of $n$ on \gams\ for any of the parameters, not even when measuring only $\Npol=3$ angles spaced by 60 degrees, indicating that using only 3 angles does not carry a precision penalty.

\begin{figure*}
	\centering
	\includegraphics[width=\textwidth]{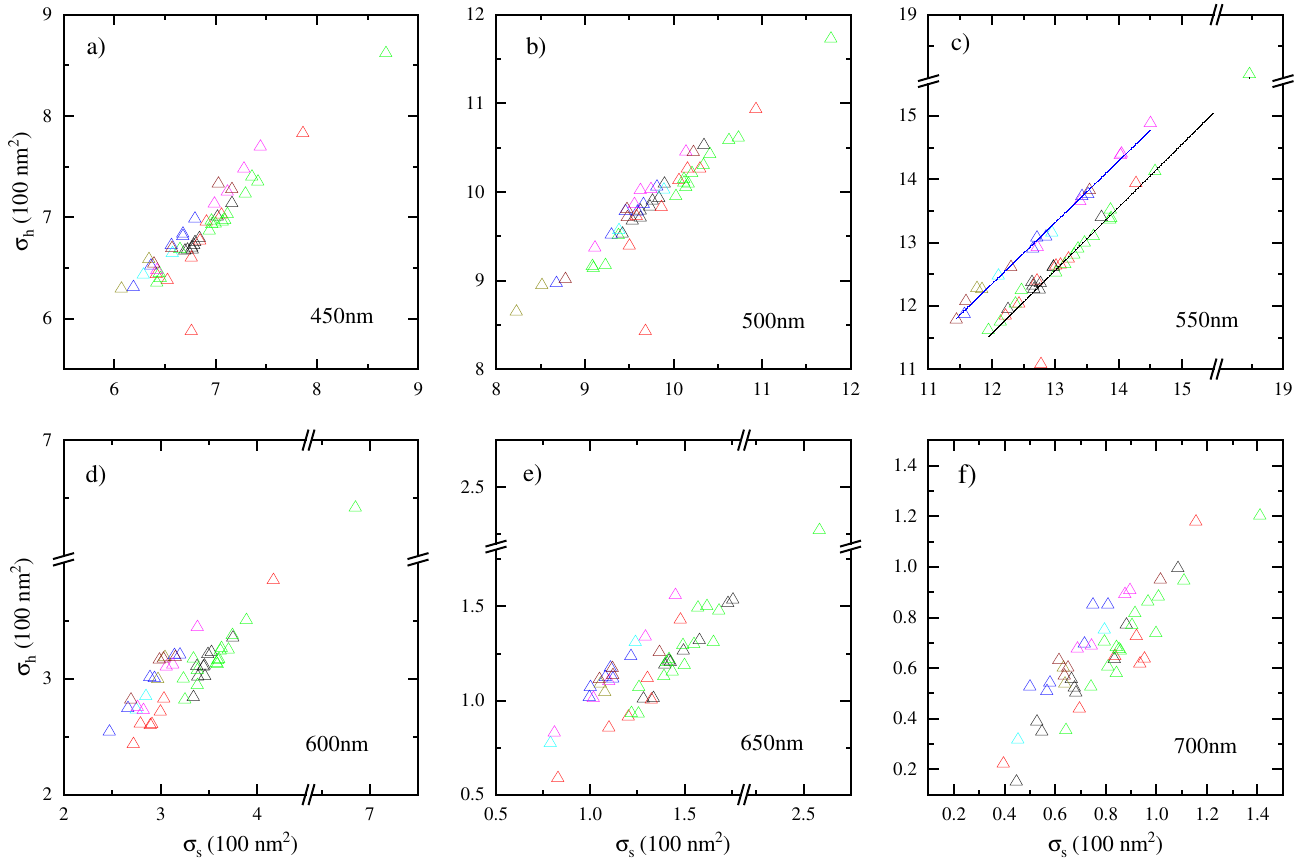}
	\caption{Extinction cross-sections, $\sigh$, measured using the larger analysis radius, \rh\, as a function of the cross-sections, $\sigs$, measured using the standard analysis radius, \rs. Panels correspond to different filter center wavelengths in the experiment as (a) $\Lambda = \cone$, (b) $\Lambda = \ctwo$, (c) $\Lambda = \cthr$, (d) $\Lambda = \cfour$, (e) $\Lambda = \cfive$, (f) $\Lambda = \csix$. Symbol colors (black, red, green, blue, cyan, magenta, dark yellow, dark red) indicate data associated to the 8 fields of view examined in the experiment. The linear fits observed in panel (c) serve to guide the eye along two clearly separated groups of datapoints.}
	\label{fig:SigComp}
\end{figure*}

\section{Influence of integration radius}\label{sec:sigcomp}

The integration radius, \ri, affects the measurement of \siex\ in wide-field extinction microscopy. For simplicity, let us first consider the more straightforward case of darkfield microscopy, where scattering by a NP results in an intensity point spread function (PSF) in the image following ideally the two-dimensional (2D) Airy distribution; in reality there will be aberrations arising from the specific optics and polarisation dependent effects. The envelope of this distribution decays asymptotically to zero with the distance from the central peak, meaning that at any finite radius some of the scattered power is not captured. One must then correct the measured integral by a factor, for example based on the known fraction of the total integral. 
By contrast, in extinction we measure a coherent signal, given by the interference of the scattered light with the incident light in forward direction, an effect also called the optical theorem. This interference results in a more extended PSF determined by additional factors, specifically the in-plane coherence length of the illumination given by the numerical aperture of the condenser, and the defocus. We have shown previously\cite{PayneSPIE19} that for suitable choice of illumination \NA\ one can achieve both a minimised fringing and therefore minimally extended PSFs, increasing the maximum possible area density of NPs in a sample, and a finite integration radius at which the extinction is fully measured.

\begin{figure*}[t]
	\centering
	\includegraphics[width=\textwidth]{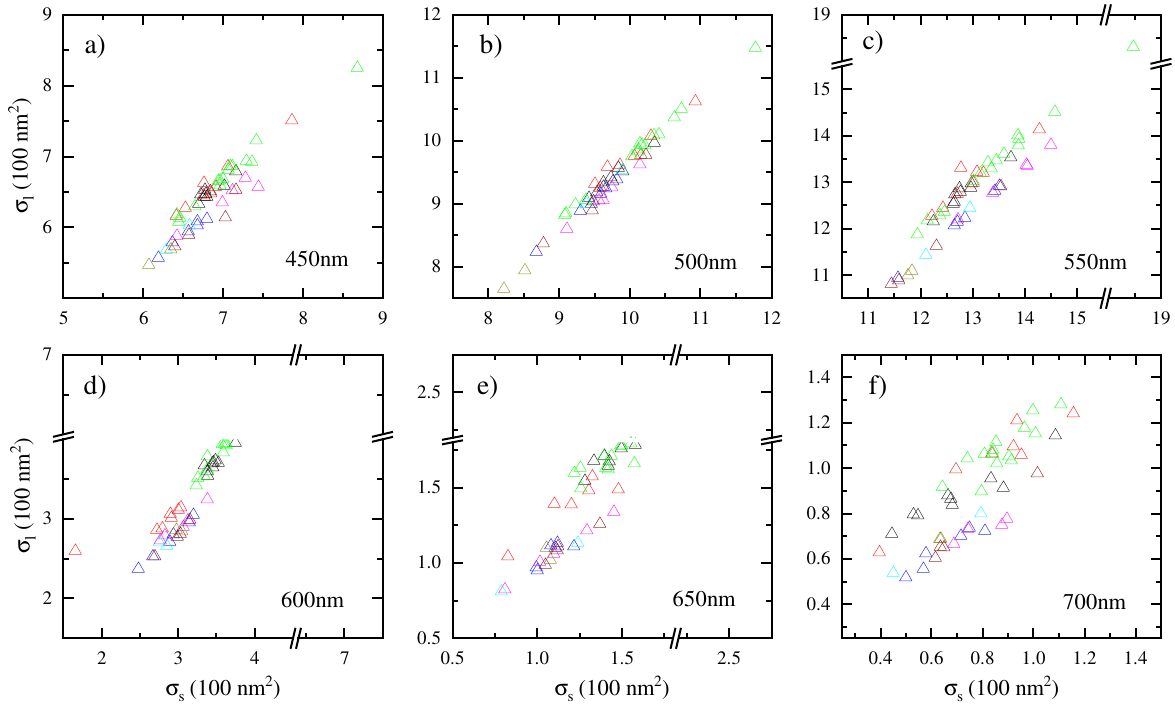}
	\caption{As for \Fig{fig:SigComp}, but using the smaller analysis radius, \rl.}
	\label{fig:SigComplow}
\end{figure*}

In this work, we chose to use the maximum opening of the condenser ($\NA=1.34$) to maximised the out-of-plane component of the field at the sample, which we employ using radially polarised light to probe the out-of-plane shape of the NPs. However, using the maximum condenser \NA\ leads to more extended fringing\cite{PayneSPIE19} in the extinction PSF, which decreases the maximum area density of NPs without spatially overlapping signals, and results in a measured extinction converging more slowly with integration radius. The latter point is the more important one for this work, since the sample area density of NPs was low. We typically use an integration radius, $\rs = 3\lambda / 2\NA$, which collects about $94\%$ of the extinguished power (dependent on condenser/objective) for non-optimal \NA\ selection. For the optimal conditions mentioned above, \rs\ would collect the full extinction, as the fringing at larger radii has compensating positive and negative contributions. Now, the discussed conditions are true assuming the particles are optimally focused, which for a purely absorbing NP can be considered to be the scenario when the peak extinction of the PSF is largest. When the focus is not optimal, the PSF spreads out, as the scattering changes its phase from being in-quadrature to the incident light, and thus not creating intensity changes, to being partly in or out of phase, creating bright and dark fringes (an effect which can be used to get contrast from non-absorbing scatters, popular in electron microscopy). In our data, a few fields of view (FOV) at a few $\Lambda$ exhibit larger PSFs (10\%-20\% larger radius to first PSF minimum) compared to others, which should not be the case if all FOV are optimally focused. This implies that one should use $\ri>\rs$ to counter the defocus effects in the analysis of \siex.

\begin{table}
	\small
	\caption{\ (IQR$\{\sigma_\mathrm{i}\}$-IQR$\{\sigs\})/\nm^2$ as a function of $\Lambda$, with subscript $i\in{(l,h)}$, indicating which measurement radius, \rl\ or \rh, was used, and the IQR calculated over all $N=51$ NPs from all 8 FOVs.} \vspace{3pt}
	\label{tab:IQR}
	\begin{tabular*}{0.48\textwidth}{@{\extracolsep{\fill}}lllllll}
		\hline
		i & 450\,nm & 500\,nm & 550\,nm& 600\,nm & 650\,nm & 700\,nm\\
		\hline
		h & -5.47 & -6.99 & 3.69 & -23.02 & -10.14 & -2.21\\
		\hline
		s & 7.49 & 11.28 & 6.10 & 33.00 & 21.67 & 10.74\\
		\hline
	\end{tabular*}
\end{table}

\begin{figure*}
	\centering
	\includegraphics[width = \textwidth]{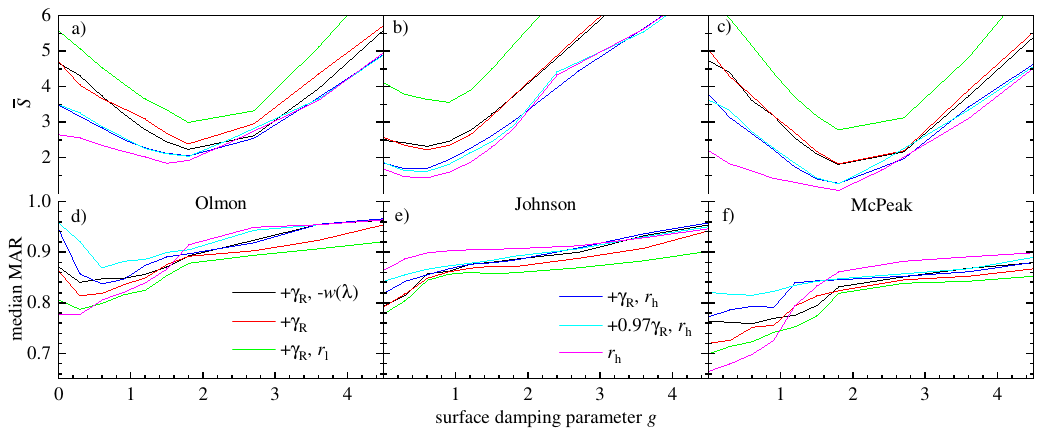}
	\caption{Median error \Sbar\ and median MAR of ON analysis as function of surface damping parameter $g$, for permittivity datasets as indicated. Results for various analysis settings are shown, labelled as follows: +\gamr, +$0.97\,\gamr$: inclusion of radially polarised measurements, unscaled or scaled by a factor of $0.97$, respectively; -\wl: no correction for the lamp spectrum and camera sensitivity; \rl, \rh: extinction data measured using the integration radius \rl\ or \rh, respectively.}
	\label{fig:Sbar_SM}
\end{figure*}

\begin{figure*}
	\centering
	\includegraphics[width=\textwidth]{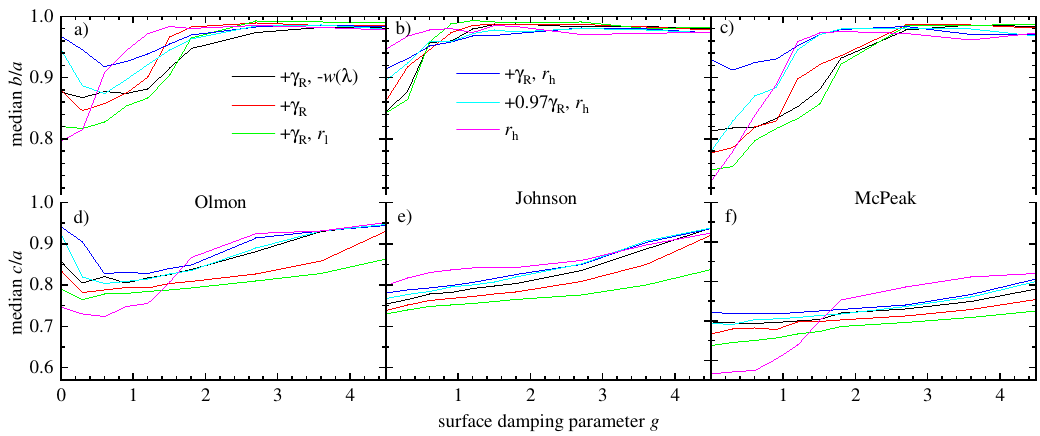}
	\caption{As \Fig{fig:Sbar_SM}, but for the median of $b/a$ and $c/a$.}
	\label{fig:Ratios_SM}
\end{figure*}

In \Fig{fig:SigComp}, we compare the polarisation-averaged \siexl\ measured using \rs, denoted as \sigs, and $\rh = 1.2\,\rs$, denoted as \sigh. We do the same in \Fig{fig:SigComplow}, but for $\rl = 0.8\,\rs$. For both figures, symbol color indicates measurements from a single FOV. In \Tab{tab:IQR} we see the difference between the interquartile ranges (IQR) of \sigs\ and \sigh\ or \sigl. Negative values across most FOV show that using \rh\ yields a smaller spread of the data, compared to \rs. The opposite is observed when using \rl\ where all values of the difference in IQR are positive, indicating an increase in the spread of data compared to \rs. For channels $\Lambda\in\{450,\,500\}\,\nm$ we can see that most of the data in \Fig{fig:SigComp} appears as a single group, specifically in \sigh, while for $\Lambda>500\,\nm$ two groups of data are visible, with black, red, and green triangles form the lower group, separating further from the rest for $\Lambda = 550\,\nm$ and $\Lambda = 650\,\nm$. For $\Lambda = 700\,\nm$ noise dominates. Note, the lines of best fit, added for $\Lambda = 550\,\nm$, indicate that certain FOVs are grouped together. We attribute this apparently wavelength-and FOV-dependent discrepancy to a slight defocus of the corresponding data. As might be expected from this attribution, the effect is more pronounced for \sigl\ in \Fig{fig:SigComplow}, since less of the signal is integrated, such that a defocus in certain FOV would result in a greater deviation compared to the well-focused cases. One can easily observe that while the same wavelength-dependent separation of FOV is apparent, certain FOV (black, grey, red triangles) form an upper group already by $\Lambda=450\,\nm$. A keen-eyed reader may note that this is the inverse behavior of \sigh, where these same FOV formed the upper group. However, this is an effect of the plotting; \sigs\ is plotted along the $x$-axis in both figures, however in \Fig{fig:SigComp}, \sigs\ yields the smaller values of the two, while in \Fig{fig:SigComplow}, \sigs\ yields the larger values of the two.

Note, the data were taken with a sample reference shift of $s = 1550\,\nm$. For our experimental setup and $\Lambda=[\cone,\,\ctwo,\,\cthr,\,\cfour,\,\cfive,\,\csix]$, we have $2\rs = [931,\,1035,\,1138,\,1241,\,1345,\,1448]\,\nm$, and $2\rh = [1117,\,1241,\,1366,\,1490,\,1614,\,1738]\,\nm$. Already for $\Lambda = [\cfive,\,\csix]$, $2\rh>s$, which is not ideal due to partial integration of signal from the dark peak into that of the bright and vice versa, reducing the measured \siex. We see that using \rh\ does lead to a lower IQR at all $\Lambda$, except for $\Lambda = \cthr$, where the defocus apparently was too large to be fully compensated. However the remaining discrepancy between the medians of the upper (five FOV) and lower (three FOV) groups of results for $\Lambda = \cthr$ is still only about $4\%$ using \rh.

\begin{figure*}
	\centering
	\includegraphics[width=\textwidth]{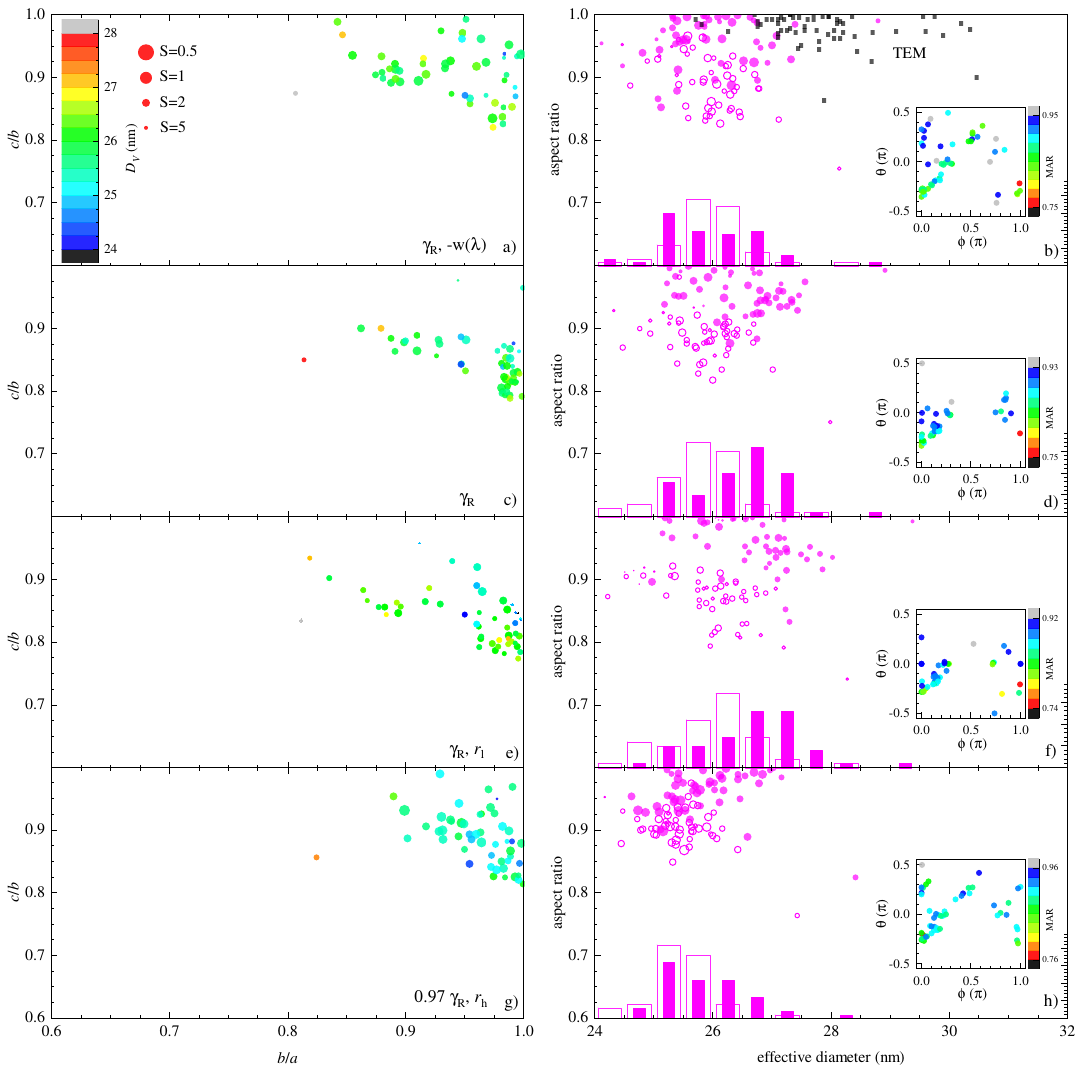}
	\caption{
		Morphometric results for $N=51$ nominally $D=30\,$nm UGNSs, for OL with $g=1.8$, similar to \Fig{fig:30nmFitPlot}, but for different analysis settings. Left: fitted aspect ratios $c/b$ versus $b/a$. The symbol size indicates the error of the fit using $(1+S)^{-1}$, as given in a), while the colour indicates the volume-equivalent diameter, \DV, as given in a). Right: MAR versus \DV\ (empty circles) and PAR versus \DP\ (filled circles), symbol sizes as in a) . In b) the TEM data PAR versus \DP\ for the $N=51$ UGNS is shown for reference. The histograms of \DV\ (empty) and \DP\ (filled) are also shown. The insets show the fitted angles, $\theta$ versus $\phi$, with a symbol colour indicating the MAR as given in the scale. Using the designations introduced in \Fig{fig:Sbar_SM}, panels (a,b) refer to +\gamr,-\wl; (c,d) to +\gamr; (e,f) to +\gamr,\rl; and (g,h) to +0.97\gamr,\rh.}
	\label{fig:multi}
\end{figure*}

Overall, assuming a fixed $s$ and sample density, this examination indicates that in future, one could optimize the illumination \NA\ to reduce \ri\ to the minimum necessary to collect the full extinction, and additionally reduce cross-talk between nearby NPs. Generally one could simply increase $s$ and use a lower sample density. Furthermore, one can improve the focussing procedure or use automated focusing techniques. This work elucidates how the accuracy of extinction measurements within an experiment can affect the spread of retrieved sizes, beyond the photon- and sample-induced measurement noise. 



\begin{figure*}
	\centering
	\includegraphics[width=\textwidth]{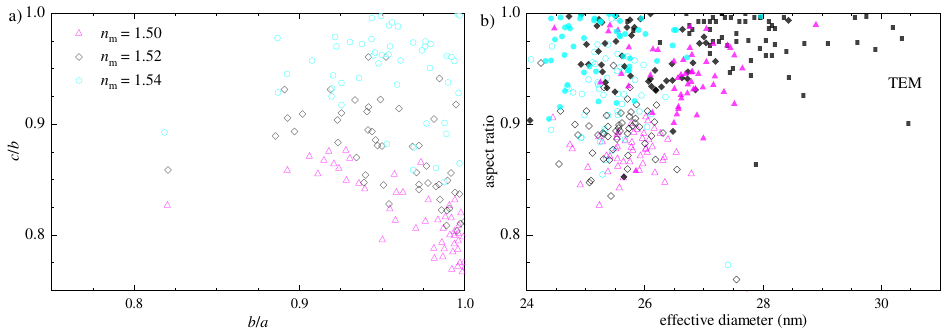}
	\caption{Morphometric results for $N=51$ nominally $D=30\,$nm UGNSs, for OL with $g=1.8$ similar to \Fig{fig:30nmFitPlot}, but for varying medium index \nmed. a) fitted aspect ratios $c/b$ versus $b/a$. b) MAR versus \DV\ (empty symbols), and PAR versus \DP\ (filled symbols).  The black filled squares indicate TEM measurements PAR versus \DP. Results are shown for the nominal index $\nmed = 1.52$ (black diamonds), a lower index $\nmed = 1.50$ (magenta triangles), and a higher index $\nmed = 1.54$ (cyan circles).}
	\label{fig:FitvsIndex}
\end{figure*}

\section{Effect of corrections on retrieved NP morphometry}\label{sec:fitcomp}

In \Fig{fig:Sbar_SM} and \ref{fig:Ratios_SM}, we explore how the various corrections and minor improvements to the extinction and shape retrieval analyses affects the outcome of the ON. Specifically, we look at the effects of inclusion of lamp spectrum and camera sensitivity, use of different \ri\ in the extinction analysis, inclusion of \siex(\gamr), and a correction of \siex(\gamr) by a factor $0.97$, compensating the difference between $\siex(\gamr)$ and $\siex(\gamp)$ as discussed in the \Sec{sec:stats} and observed in \Fig{fig:sigma_plots}. We find (see \Fig{fig:Sbar_SM}a-c) that the dominant effect on \Sbar\ are offsets, while the value of $g$ for minimum \Sbar\ is less affected. The largest \Sbar\ is found using $\rl$, as expected from the larger defocussing artefacts. We find that when \siex(\gamr) is not included, \Sbar\ reaches its lowest value across all cases, as expected by the reduced constraints on the fit, see also \Sec{sec:morph}. Including \siex(\gamr), the lowest \Sbar\ is found using $0.97\siex(\gamr)$, including \wl, and using \rh\ in the extinction analysis. Across the analysis conditions we find \Sbar\ is smallest for McPeak, and largest for Olmon.

In \Fig{fig:Sbar_SM}(d-f) and \Fig{fig:Ratios_SM} we show the dependence of the aspect ratios on $g$ for the different permittivity datasets and analysis conditions. A considerably stronger dependence on $g$ is observed for \brat\ than for \crat, with \brat\ obtaining close to its maximal value typically for \Sbar\ at its minimum, and plateauing for increasing $g$. This can be understood by the broadening of the resonance, increasing the cross-sections at wavelengths longer than the LSPR, which otherwise requires non-spherical shapes shifting the LSPR to the red.

We show the distributions of retrieved shapes and sizes for the different analysis conditions in \Fig{fig:multi}. The condition +0.97\gamr,+\rh\ (panels g,h) provides the smallest spread and the roundest particles, closest to the shape seen in TEM of all conditions apart from those using $\ep^s(\lambda)$ seen in \Sec{sec:morph} and \Fig{fig:FitPlotShift}. At the same time, it provides the smallest \DV, about 2\,nm lower than TEM. This points towards a systematic error of the model. As discussed in \Sec{sec:morph} and \Sec{sec:Conc}, the most likely source for this is retardation effects leading to a reduction of the cross-section, broadening by radiative damping, and red-shift compared to the static approximation used in our model. 


\section{Effect of medium refractive index on retrieved NP morphometry}\label{sec:fitindex}

The effects of deviations of the refractive index of the NP environment from its nominal value $\nmed = 1.52$ on the morphometric retrieval are shown in \Fig{fig:FitvsIndex} for OL with $g=1.8$, with other analysis conditions as for the results seen in \Fig{fig:30nmFitPlot}. We use changes of $\pm 0.02$, which fully cover the range that might be expected from spectral dispersion in glass or silicone oil over the probed wavelength range, or batch-to-batch material variations. We find that the dominant effect is on the retrieved shape, specifically on  the AR $c/b$ which changes by about twice the index change.
Such a behaviour is expected from the red-shift of the LSPR with increasing \nmed, which is compensated by a reduced asymmetry-induced red-shift. These results indicate that the medium refractive index accuracy needs to be better than half the required aspect ratio accuracy. A slight change of the retrieved size is also seen, by about $\mp 0.25$\,nm for $\pm 0.02$ index change.

\section{Effect of retardation on retrieved NP morphometry}\label{sec:fitmodel}

\begin{figure}[b]
	\centering
	\includegraphics[width=\columnwidth]{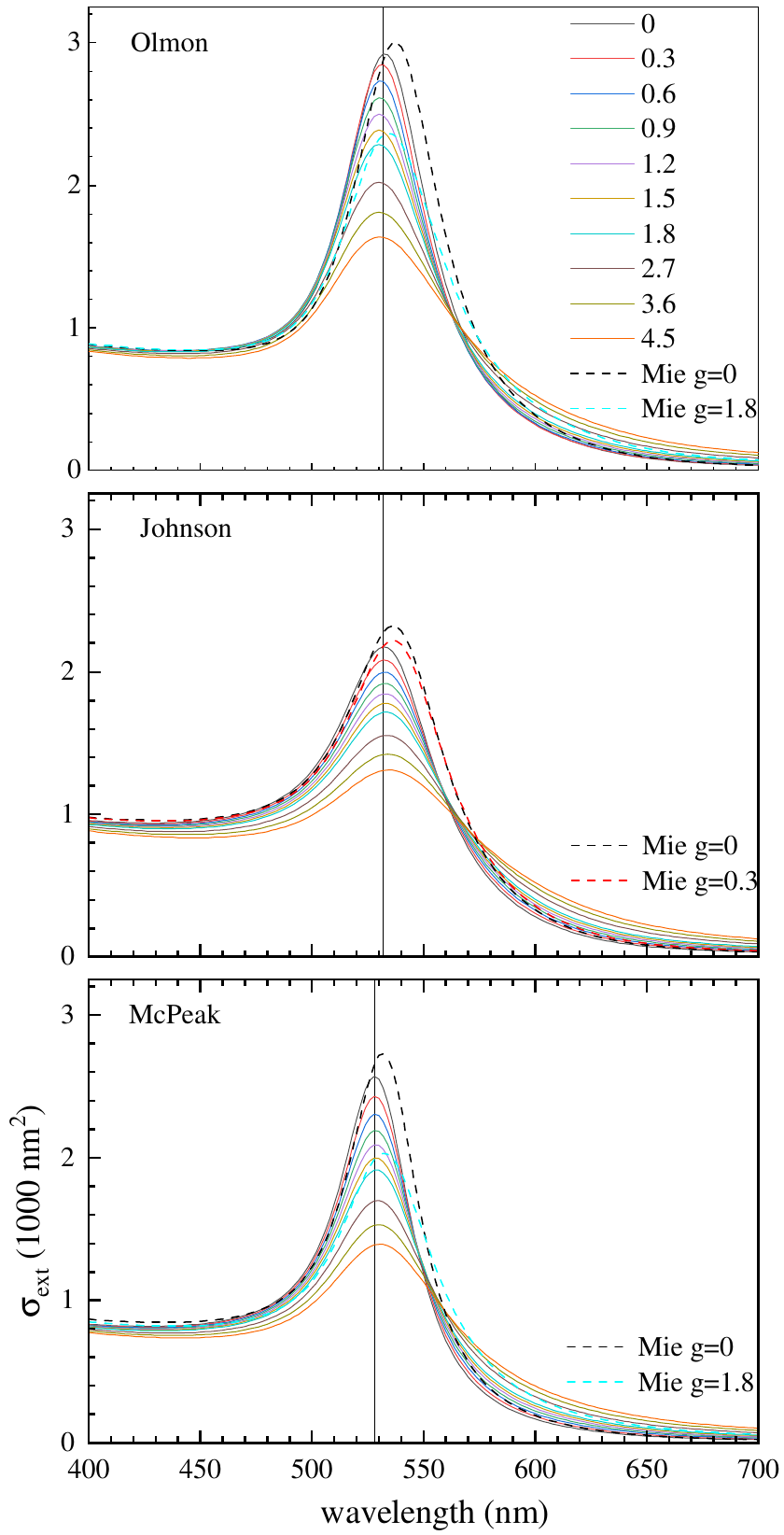}
	\caption{Solid curves : $\siex(\lambda)$ for a spherical NP of $27.94\,\nm$ diameter calculated for a range of $g$ as indicated and each of the three permittivity datasets described in the main text using the Rayleigh-Gans theory. Dotted curves : as above, but calculated using Mie theory at $g=0$ and the $g$ at which minimum \Sbar\ was obtained in the morphometric retrieval for each of the three permittivities ($g=(1.8, 0.3, 1.8)$ for OL, JC, and MP, respectively. Vertical lines indicate the peak position in $\siex(\lambda)$ for $g=0$ for the Rayleigh-Gans theory.}
	\label{fig:extspec}
\end{figure}


\begin{figure}[b]
	\centering
	\includegraphics[width=\columnwidth]{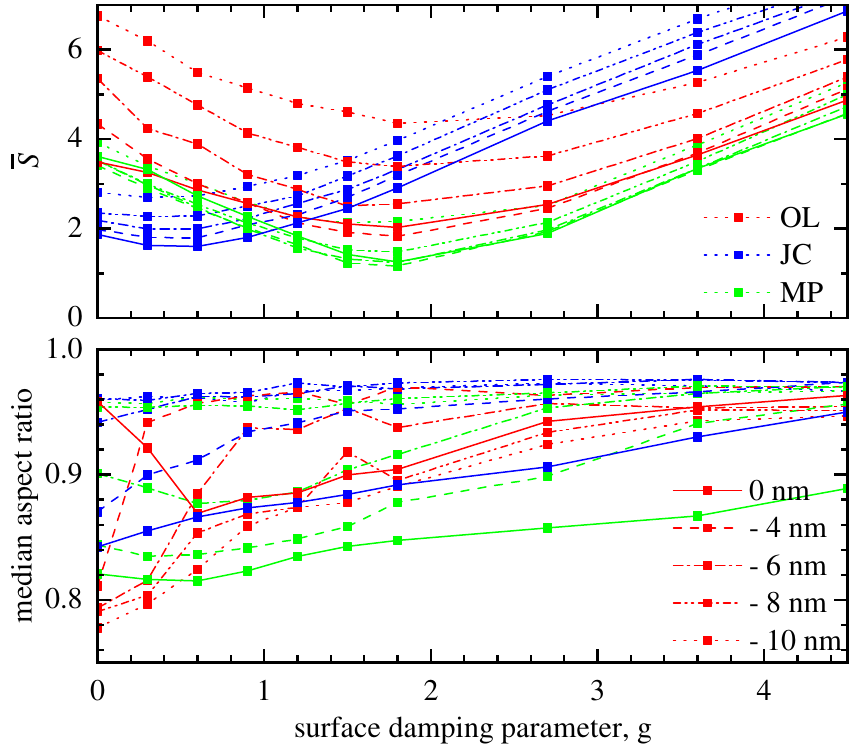}
	\caption{\Sbar\ and median MAR for $N=51$ UGNS of nominal $D=30\,$nm for varying surface damping parameter $g$ as in \Fig{fig:ErrorOverview}, but using $\ep^s(\lambda)$ for different $\ep$ datasets indicated by color and wavelength shifts $\delta\lambda$ indicated by line type, as labelled. }
	\label{fig:shiftcomp}
\end{figure}

Here, we consider if the few percent difference in aspect ratio and size between the retrieved morphometries and the TEM-measurements is in part attributable to the fact that the dipole approximation does not take into account retardation effects. Specifically, in \Fig{fig:extspec}, we show the extinction spectra for a spherical NP of $27.94\,\nm$ diameter plotted for all $g$ used in the analysis for each of the three permittivity datasets in the main text calculated using the Rayleigh-Gans theory. These are compared with calculations using Mie theory (dotted lines) at $g=0$ and $g$ yielding the minimum \Sbar\ for each of the three permittivity datasets. Importantly, we see that there is a small but noticeable red-shift in the peak of spectra calculated using Mie theory which includes retardation. Such a redshift can in the Rayleigh-Gans theory only be achieved by increasing the shape asymmetry. To account for the red-shift from retardation without introducing a new model into the analysis, we shift the permittivity data by $\delta\lambda \in {\{-4,-6,-8,-10\}}$\,nm, introducing $\ep^s(\lambda) = \ep(\lambda+\delta\lambda)$. We then perform the ON analysis using $\ep^s$. To ensure we are not limited by the discussed small discrepancy between the linear and radial measurements of \siex, we performed this additional analysis using the $0.97\times$ correction to the \siex(\gamr)\ data.

We examine the resulting dependence of \Sbar\ and the median MAR and PAR on $g$ for the different $\delta\lambda$ in \Fig{fig:shiftcomp}. The JC permittivity is the only one where \Sbar\ does not decrease for some $\delta\lambda<0$. Meanwhile, OL and MP both show a minimum of \Sbar\ for $\delta\lambda = -4\,\nm$, before the error begins increasing past the original $\delta\lambda = 0$ case. MP shows a less clear dependence of MAR on $g$ for the different $\delta\lambda$ compared to OL, but in all permittivities, increased $\delta\lambda$ shows increased average roundness of the retrieved morphometries.


%

\end{document}